\documentclass[manuscript,screen]{acmart}

\usepackage{graphicx}
\usepackage[export]{adjustbox}

\AtBeginDocument{%
  \providecommand\BibTeX{{%
    \normalfont B\kern-0.5em{\scshape i\kern-0.25em b}\kern-0.8em\TeX}}}

\acmPrice{15.00}

\acmSubmissionID{3340}

\copyrightyear{2024}
\acmYear{2024}
\setcopyright{acmlicensed}\acmConference[CHI '24]{Proceedings of the CHI Conference on Human Factors in Computing Systems}{May 11--16, 2024}{Honolulu, HI, USA}
\acmBooktitle{Proceedings of the CHI Conference on Human Factors in Computing Systems (CHI '24), May 11--16, 2024, Honolulu, HI, USA}
\acmDOI{10.1145/3613904.3642165}
\acmISBN{979-8-4007-0330-0/24/05}

\begin{document}

\title[IntentTuner]{\textit{IntentTuner}: An Interactive Framework for Integrating Human Intentions in Fine-tuning Text-to-Image Generative Models}

\author{Xingchen Zeng}
\email{xingchen.zeng@outlook.com}
\affiliation{
  \institution{The Hong Kong University of Science and Technology (Guangzhou)}
  \city{Guangzhou}
  \country{China}
}
\author{Ziyao Gao}
\email{gaoziyao@hkust-gz.edu.cn}
\affiliation{%
  \institution{The Hong Kong University of Science and Technology (Guangzhou)}
  \city{Guangzhou}
  \country{China}
}
\author{Yilin Ye}
\email{yyebd@ust.hk}
\affiliation{%
  \institution{The Hong Kong University of Science and Technology (Guangzhou)}
  \city{Guangzhou}
  \country{China}
}
\affiliation{
  \institution{The Hong Kong University of Science and Technology}
  \city{Hong Kong SAR}
  \country{China}
}
\author{Wei Zeng}
\email{weizeng@hkust-gz.edu.cn}
\authornote{Wei Zeng is the corresponding author}
\affiliation{%
  \institution{The Hong Kong University of Science and Technology (Guangzhou)}
  \city{Guangzhou}
  \country{China}
}
\affiliation{
  \institution{The Hong Kong University of Science and Technology}
  \city{Hong Kong SAR}
  \country{China}
}
\renewcommand{\shortauthors}{Zeng et al.}

\newcommand{\eg}{\textit{e.g.}}
\newcommand{\etc}{\textit{etc}}
\newcommand{\ie}{\textit{i.e.}}
\newcommand{\etal}{\textit{et al.}}
\newcommand{\red}[1]{{\color{red}{#1}}}
\newcommand{\revise}[1]{{\color{black}{#1}}}


\begin{abstract}
Fine-tuning facilitates the adaptation of text-to-image generative models to novel concepts (\eg, styles and portraits), empowering users to forge creatively customized content. 
Recent efforts on fine-tuning focus on reducing training data and lightening computation overload but neglect alignment with user intentions, particularly in manual curation of multi-modal training data and intent-oriented evaluation.
Informed by a formative study with fine-tuning practitioners for comprehending user intentions, we propose \emph{IntentTuner}, an interactive framework that intelligently incorporates human intentions throughout each phase of the fine-tuning workflow. 
\emph{IntentTuner} enables users to articulate training intentions with imagery exemplars and
textual descriptions, automatically converting them into effective data augmentation strategies. Furthermore, \emph{IntentTuner} introduces novel metrics to measure user intent alignment, allowing intent-aware monitoring and evaluation of model training. Application exemplars and user studies demonstrate that \emph{IntentTuner} streamlines fine-tuning, reducing cognitive effort and yielding superior models compared to the common baseline tool.
\end{abstract}





\maketitle

\section{Introduction}
Recent advancements in pre-trained text-to-image generative models, such as Stable Diffusion~\cite{rombach2022high} and DALL-E-2~\cite{ramesh2022hierarchical}, have facilitated the generation of high-quality images from natural language descriptions (\ie, prompts)~\cite{saharia2022photorealistic}. 
These technologies have shown promise in augmenting creative processes across various fields (\eg, news illustrations~\cite{liu2022opal}, fashion design~\cite{wu2023styleme}, and webtoon~\cite{ko2022we}).
\revise{However, pre-trained models often fall short in catering to users' diverse and domain-specific demands, particularly when dealing with concepts not included in the training data.}
Consequently, users increasingly need to tailor text-to-image generation to their unique requirements on design concepts such as styles and clothing or facial features in portraits. 
Such demand has spurred the community of artificial intelligence (AI) art and AI design users to learn and adopt \textbf{\textit{fine-tuning}}, a powerful technique that enables pre-trained models to learn new concepts with a small number of additional training examples reflecting users' desired outcomes. 
This practice has given rise to burgeoning online communities for sharing models fine-tuned by users, such as "Civitai"~\cite{civitai2022} and "Liblib AI"~\cite{liblib2022}.
\revise{Furthermore, fine-tuning tutorials targeting non-expert users are receiving considerable attention, expanding the user base beyond AI experts to include artists, designers, and novice users.}


\begin{figure}[t]
\centering
 \includegraphics[width=0.995\textwidth]{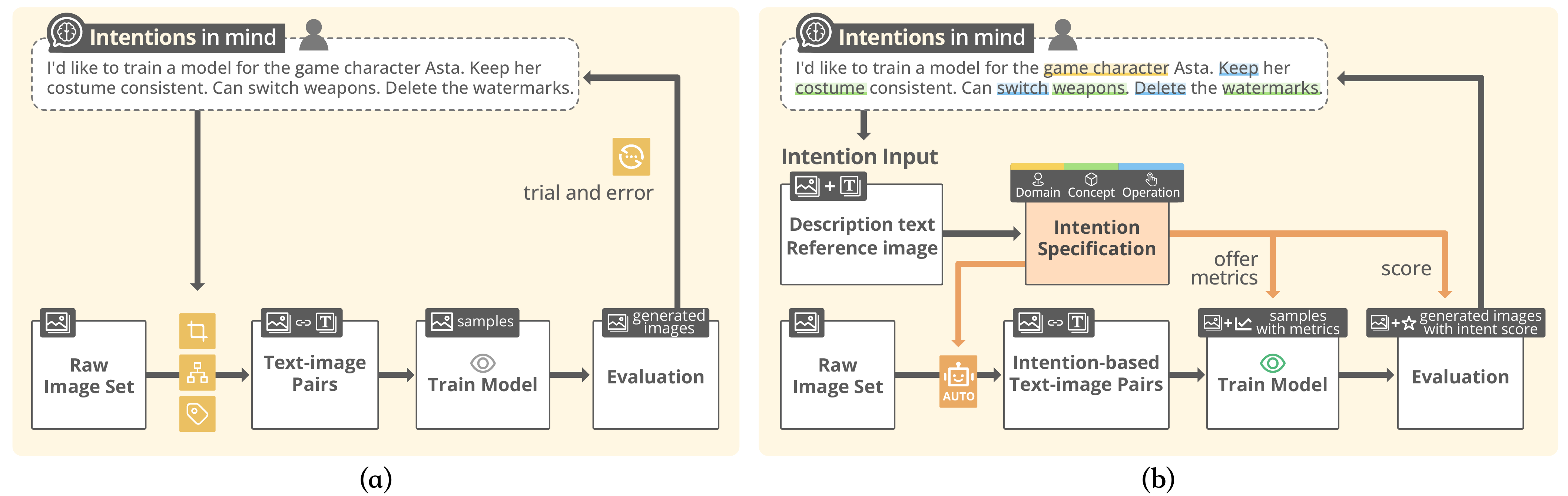}
 \vspace{-4mm}
\caption{
\textbf{Comparison of pipelines of general and our intent-aligned fine-tuning framework.} 
(a) General pipeline. Most users rely on a trial and error process to check whether the system properly understands their intents, where they manually preprocess the training images, 
such as cropping {\includegraphics[height=1.1em,valign=c]{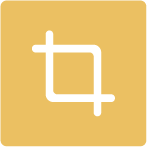}}, categorizing {\includegraphics[height=1.1em,valign=c]{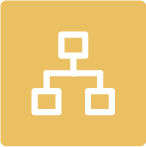}} and tagging {\includegraphics[height=1.1em,valign=c]{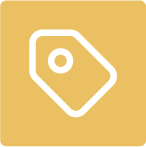}}, and observe the generated images. 
(b) Our pipeline. \textit{IntentTuner} allows users to efficiently articulate their intents to automatically steer important milestones of the fine-tuning, including data augmentation, training monitoring, and evaluation.
}
\vspace{-4mm}
\label{fig:workflow_compare}
\end{figure}

Existing research has primarily focused on developing efficient fine-tuning methods~\cite{ruiz2023dreambooth, hu2021lora, gal2022image} from the model's perspective, aiming to reduce the number of images required for a model to learn new concepts (\eg, DreamBooth~\cite{ruiz2023dreambooth}) and to lighten computational resources demands (\eg, Low-rank adaptation~\cite{hu2021lora}).
However, after selecting a specific fine-tuning method, effectively applying it remains challenging for AI novices and even for AI experts.
\revise{When fine-tuning, users often have \textbf{conceptual intents} in mind, including what features the model should learn to keep and what features should be modified or deleted, such as shown in Fig.~\ref{fig:workflow_compare} \raisebox{-.2\height}{\includegraphics[width=0.32cm]{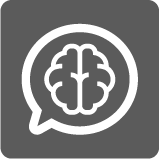}}.}
\revise{Difficulties arise when further aligning these intents with the technical steps involved in the fine-tuning process.}
\revise{Specifically, users are expected to translate their intents} into concrete data strategies (\eg, image augmentation and caption optimization) and performance assessments (\eg, model evaluation and selection), which is currently a trial-and-error process (Fig.~\ref{fig:workflow_compare} (a)).
Moreover, measuring the quality of generated images has been a challenging and ongoing research problem~\cite{shan2023glaze}, especially evaluating the alignment between the fine-tuned model and user intents~\cite{wu2023better}.
Recently, the community has developed interactive fine-tuning tools (\eg, Koyhass~\cite{kohyass2022}), offering detailed training settings, automated image captioning, and metrics visualization to aid in monitoring the training process.
However, the tools do not address the challenges above for intent alignment and exhibit an "engineering mindset," only allowing users to control low-level settings and observe hard-to-interpret predefined metrics without an explicit connection to user intents.
Moreover, these fine-tuning systems are not integrated with the generation interface, requiring users to load the fine-tuned models on a separate UI (\eg, SD WebUI~\cite{sdui2022}) to try out generation.
Thus, existing tools are insufficient for improving fine-tuning results, reducing the trial-and-error workload, and increasing the accessibility to a broad user community.

To tackle these challenges, we present \textit{IntentTuner}, an interactive framework for integrating user intents into fine-tuning text-to-image generative models with threefold considerations:
1) understanding user intents via natural descriptions and interactions; 
2) efficiently translating user intents into intent-aligned data strategies; 
and 3) monitoring and evaluating the training pipeline in an intent-aligned manner.
The design of \textit{IntentTuner} draws insights from a preliminary study that delves into comprehending the practical fine-tuning workflow, the user intent structure, and the challenges users encounter (Sect.~\ref{Sec-PreliminaryStudy}).
As shown in Fig.~\ref{fig:workflow_compare} (b), our framework translates multi-modal user input into intent specifications to explicitly guide the data augmentation, training monitoring, and model evaluation (Sect.~\ref{Sec-framework}).

\revise{
To facilitate our framework, we have developed an interactive system that enables users to articulate their intents effortlessly by
focusing on specific concepts of target effect images with visual interaction and clarifying their intents with natural language (Sect.~\ref{ssec:understanding-user-intent}).
These inputs are then mapped to clear intent specifications, including \emph{\textbf{Keep}}, \emph{\textbf{Modify}}, and \emph{\textbf{Delete}} intents.
Specifically, as shown in Fig.~\ref{fig:workflow_compare} (b), 
they are organized into a hierarchy of domain-concept-operation, linking associated domain (\eg, \textit{game character}) and concepts (\eg, \textit{costume-keep}, \textit{weapons-modify}, and \textit{watermarks-delete}).
}
Based on the intent specifications and the characteristics of the training dataset, the system automatically generates strategies for data processing, basic hyperparameter setting, and model evaluation (Sect.~\ref{ssec:intent-guided-data-strategy}).
\revise{Our system subsequently supports intuitive training monitor and evaluation with intent-specific metrics combined with sample generation (Sect.~\ref{ssec:system-evaluation}).}
\revise{Overall, our framework and system} aid users in efficiently and intuitively creating and evaluating fine-tuned models, reducing cognitive load during the process, enhancing the performance of the original dataset, and improving training and iteration efficiency. 

In summary, our major contributions include:
\begin{itemize}
    \item We introduce a novel framework to intelligently integrate human intents in fine-tuning text-to-image generative models by effectively decomposing user inputs into intent-aligned data augmentation, model monitoring, and evaluation strategies.
    \item We develop an integrated system that unifies the fine-tuning and generation into a holistic interface that enables both expert and novice users to flexibly customize text-to-image generation models based on their intents expressed in multi-modal natural inputs. It simultaneously supports user-friendly monitoring and evaluation, facilitating the intuitive selection of generation models.
\end{itemize}

\noindent \revise{\textbf{Ethics statement}. All the human portrait images used in this study do not concern any celebrities or other existing human beings. 
The human portrait fine-tuning data are synthetic and for demo purposes only.
The models obtained by our experiment will not be used without consent to generate images imitating any real person.}

\section{Related Works}

\subsection{Text-to-Image Generative models and Fine-tuning}
Text-to-image generative models have showcased impressive capabilities of translating natural language description to high-quality images~\cite{ramesh2022hierarchical, saharia2022photorealistic, rombach2022high}, surpassing conventional mainstream generative adversarial networks (GANs)~\cite{gal2022stylegan, karras2019style} and auto-regressive models~\cite{ramesh2021zero, yu2022scaling}. 
Representatively, Imagen~\cite{saharia2022photorealistic} discovers that pre-trained language models are surprisingly effective at encoding text for image synthesis.
Rombach et al.~\cite{rombach2022high} introduced latent diffusion models (\ie, stable diffusion), where the forward and reverse diffusion processes happen on the latent space learned by an auto-encoder, remarkably reducing the computational cost. 
Following works bring further improvements on various down-stream tasks, such as image editing~\cite{kawar2023imagic, brooks2023instructpix2pix}, inpainting~\cite{lugmayr2022repaint, yu2023inpaint}, and style transferring~\cite{zhang2023inversion}, directly contributing to the explosive growth of the Artificial Intelligence Generated Content (AIGC) community~\cite{liblib2022, civitai2022} and real-world applications. 
These advancements have also inspired HCI researchers to explore how human users can harness generative models for AI-supported creation or human-AI co-creation~\cite{inie2023designing, verheijden2023collaborative, liu2023beyond}.
Particularly, many studies focus on developing tools~\cite{wang2023reprompt, feng2023promptmagician, brade2023promptify} or guidelines~\cite{liu2022design} to help users adjust prompts to optimize the quality of generated images.

Nevertheless, the outputs of pre-trained diffusion models are inherently constrained by their training corpus, sometimes even directly copying the data~\cite{somepalli2023diffusion}.
Although techniques like prompt engineering, image in-painting~\cite{lugmayr2022repaint} and  editing~\cite{gal2022stylegan,brooks2023instructpix2pix}, and other image modifications~\cite{zhang2023adding} can help users iteratively prompt the models to refine the generation results. 
These methods still rely on concepts already learned by pre-trained models.
Consequently, no matter how much effort is invested in prompt design, they still fall short when confronted with "unseen" concepts spanning from abstract styles to specific content types.
To mitigate this issue,  researchers have introduced various fine-tuning techniques aimed at instructing pre-trained models about novel concepts, such as Textual Inversion~\cite{gal2022image}, DreamBooth~\cite{ruiz2023dreambooth}, and Low-rank Adaptation (LoRA)~\cite{hu2021lora}. 
Out of these approaches, LoRA has gained substantial traction within the community for significantly reducing the number of trainable parameters and thus minimizing the need for GPU memory.  
Our work is designed to support a wide range of users, thus choosing LoRA to maximize its accessibility.

However, the effectiveness of fine-tuning does not solely depend on the chosen training technique but also significantly on other crucial factors, especially the availability of high-quality training images with target concepts embedded and textual descriptions that align well with user intents.
Particularly, fine-tuning without well-aligned high-quality data leads to models lacking controllability and generating images excessively resembling patterns in the training data.
Furthermore, many models are generated during the fine-tuning process, posing a challenge in efficiently selecting the model that balances controllability and alignment with the user's intent.
Our work contributes to the automatic optimization of training data (\ie, text-image pairs) by utilizing user intents as a guiding principle and helping users pick the intent-aligned model from multiple perspectives.

\subsection{Evaluation of Text-to-Image Generation}
Evaluating the quality of text-to-image generation has been a challenging and ongoing research problem due to the subjective nature of image evaluation and the inherent gap between text and image modalities~\cite{shan2023glaze}.
\revise{
Specifically for evaluating the generation after fine-tuning, there are two aspects to consider: the model's ability to replicate target concepts and its controllability in modifying concepts using different textual prompts~\cite{gal2022image}.}
\revise{
Traditional image quality metrics like saliency scores~\cite{borji2019salient} fall short in evaluating similarities between images and establishing connections with the text modality. 
}
Inception Score~\cite{salimans2016improved} and Fréchet Inception Distance~\cite{heusel2017gans} are commonly used to assess generative models by measuring distributional differences between generated images and training images.
However, they cannot evaluate either single-image generations or text-image consistency.
To evaluate individual generated images based on a prompt, previous studies~\cite{yu2022scaling,shan2023glaze} often employ metrics based on Contrastive Language-Image Pre-Training (CLIP)~\cite{radford2021learning}, which compute text-image consistency by cosine similarity between text and image embeddings in the joint representation space.
To better align with human preferences, researchers explored fine-tuned CLIP using datasets of human ratings on images created from identical prompts~\cite{wu2023better, xu2023imagereward, dinh2021tise}. 
They further utilized scores predicted by the fine-tuned CLIP to approximate human assessment.

\revise{However, these works have focused on evaluating the overall images.
Although those board metrics and other narrow metrics (\eg, color harmonious~\cite{cohen2006color}) are useful, the granularity of practical user intents is more moderate, \ie, the alignment between the generated images and users' specific intended concepts (\eg, the hair color of a portrait and the face similarity).
Moreover, previous studies have not considered the controllability of the fine-tuned models in modifying fine-grained user-intended attributes, which is a critical factor in evaluating the overfitting of fine-tuning.
In our work, we consider both the replication and modification of user-intended concepts at a moderate granularity, ensuring that the outcomes align with user intents while avoiding overfitting.
}

\subsection{Intent Understanding with Large Vision-Language Models}
Understanding user intent and incorporating it into interactive systems to make them more user-friendly and intelligent is a common topic of interest in the HCI community. 
\revise{Both language and vision are powerful communication channels.}
Recently, large pre-trained language models (LLM) have shown significant ability to comprehend fuzzy text inputs, stimulating the development of natural language interfaces for user interactions~\cite{touvron2023llama, brown2020language, liang2023taskmatrix, ross2023programmer, shen2023hugginggpt}.
For example, Ross et al.~\cite{ross2023programmer} investigated the feasibility of using conversational interactions based on code and whether software engineers are open to conversing with LLM. HuggingGPT~\cite{shen2023hugginggpt} helps users match their natural language inputs with AI-assisted task requirements and directs users toward the most suitable model published in machine learning communities (\eg, Hugging Face).
\revise{Regarding the vision channel, Vinker \etal~\cite{vinker2023concept} presented an approach for decomposing user-provided exemplar images into distinct visual elements. 
This resulted in a hierarchical structure of sub-concepts, which could be combined and explored through textual inversion to generate imaginative ideas.}

In specific scenarios, \revise{language or vision alone} may be insufficient to convey complex user intents, particularly in visual or cross-modal tasks such as interactive image segmentation~\cite{kirillov2023segment, li2023semantic} and visual question answering~\cite{wu2017visual}.
To address this problem, researchers have been exploring using visual-language models ~\cite{li2023blip} to support both language and vision channels.
For example, the Segment Anything Model (SAM)~\cite{li2023semantic} allows users to combine textual and visual prompts to precisely segment fine-grained elements within an image.
SAM's visual prompts empower users to effortlessly click anywhere on the image or draw a straightforward bounding box to convey their intent for selecting semantic areas. 
With textual prompts, SAM can be guided to segment objects of interest based on their semantic properties.
This innovative approach enhances understanding of user intents and holds substantial promise for HCI applications involving visual data, such as personalized image inpainting~\cite{yu2023inpaint} and interactive image matting~\cite{yao2023matte}. 

However, as user intentions for fine-tuning mainly involve infusing new semantic concepts and aligning with the desired visual features, intent comprehension in any single modality is insufficient. 
In addition, existing multi-modal prompting models like SAM are not tailored to fine-tuning and \revise{are unclear as to how to make the vision and language channels complement each other~\cite{zhang2023comprehensive}}.
To fill the gap, our work leverages textual and visual channels to effectively enable users to articulate their fine-tuning intentions. 
By combining complementary cues from both language and vision channels, we adopt a cross-modal strategy to holistically translate user intents, leading to an intent-aligned optimization of the fine-tuning process.

\section{Preliminary Study}
\label{Sec-PreliminaryStudy}
We conducted a preliminary study (Sect.~\ref{ssec:pre_study}) to comprehend the current practice of fine-tuning text-to-image models and the pain points.
From the study, we summarized general workflow (Sect.~\ref{sssec:pre_workflow}), intentions (Sect.~\ref{sssec:pre_intention}), and challenges users encounter (Sect.~\ref{sssec:pre_challenge}) in the fine-tuning process.
Based on the findings, we formulated design goals to more effectively integrate human intentions into the fine-tuning of text-to-image models (Sect.~\ref{ssec:pre_goal}).

\subsection{Study Design}\label{ssec:pre_study}
\noindent \textbf{Participants.} To ensure our tool is widely accessible to users within the model fine-tuning community, we conducted observational tasks and semi-structured interviews with experts from diverse backgrounds: 
an enthusiast model trainer \revise{majoring in e-commerce} (P1, Male, Age: 23) that fine-tunes and shares models online but has limited knowledge in technical principles of fine-tuning, and struggles to obtain high-quality datasets and models consistently; 
two intermediate model trainers \revise{with majors in computer science and industrial design} (P2-P3, Males, Ages: 23-27) that fine-tune models for academic research and have studied the impact of various settings on training outcomes and possess high-quality datasets and models in specific domains;
and a professional model trainer \revise{with a background in illustration} (P4, Female, Age: 25) that fine-tunes models for commercial use, and has higher demands on result stability and controllability and acquired a comprehensive understanding of technical principles of model fine-tuning.


\noindent \textbf{Procedure.}
We initially collected demographic information from the participants and requested them to present their previous work to confirm their expertise in fine-tuning. 
Next, each participant was asked to complete two fine-tuning tasks in two distinct domains, followed by an interview. 
In \textit{Task 1}, participants were asked to perform a complete fine-tuning process in their familiar domain using the dataset they had prepared. 
In \textit{Task 2}, they were assigned to fine-tune a model in an unfamiliar domain.
Finally, we conducted interviews with the participants. 
They were asked about their workflows when fine-tuning the text-to-image models, including adjustments to specific strategies in workflow based on their training intentions and the challenges they encountered throughout the process. 
The entire process lasted from 90 to 120 minutes.
After the interviews, we transcribed the audio recordings, and the findings are summarized below.

\subsection{Findings}

\subsubsection{General Workflow}\label{sssec:pre_workflow}

\begin{figure}[t]
\centering
 \includegraphics[width=0.895\textwidth]{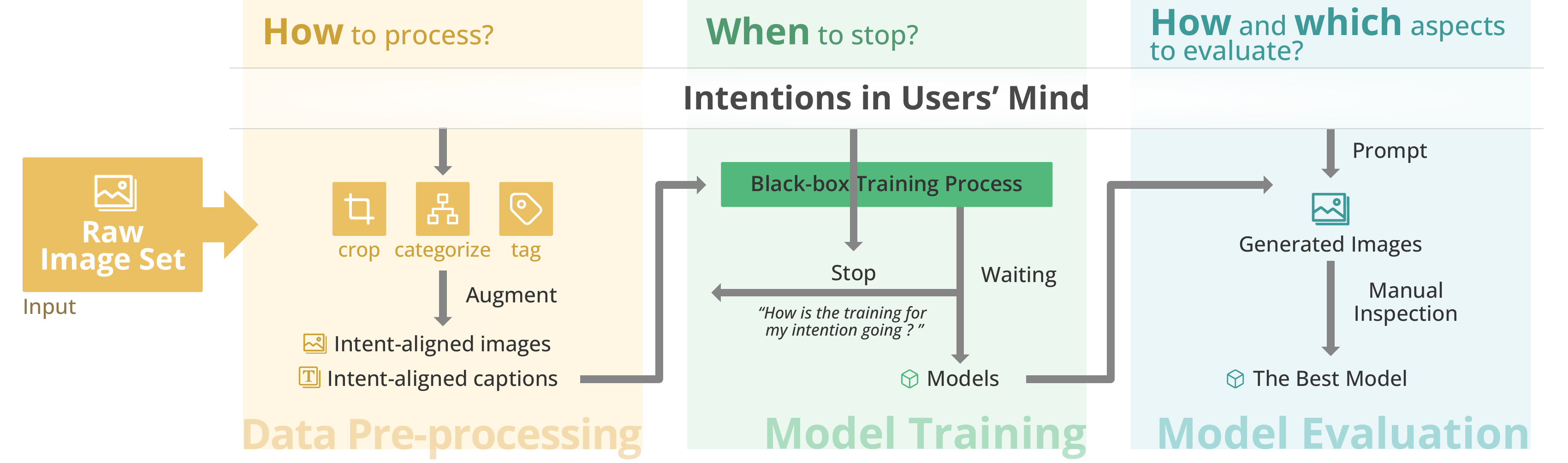}
 \vspace{-2mm}
\caption{
\textbf{General workflow.}
The input Raw Image Set is enhanced during the \textbf{Data Pre-processing} phase by cropping, categorizing, and tagging according to the intended requirements, producing images and captions that align with the intention. 
Next, using the processed image-caption pairs, the \textbf{Model Training} begins. 
Users can monitor the progress of model training to determine whether to continue or stop.
Finally, users generate images to conduct \textbf{Model Evaluation. }
Users manually input various prompts to test the model's performance from perspectives related to the intention. 
After manual inspection, the optimal model is selected.
}
\vspace{-4mm}
\label{fig: general workflow}
\end{figure}

As depicted in Fig.~\ref{fig: general workflow}, the general workflow of model fine-tuning encompasses three phases, including \textit{Data Pre-processing}, \textit{Model Training}, and \textit{Model Evaluation}.

\begin{itemize}
\item
\textbf{Data pre-processing.} 
In this phase, users categorize, crop, and tag the raw dataset, \revise{to enhance visual features that align with the intention meanwhile reducing undesired content.
For example, P1 manually duplicated original images and then cropped the clothing while fine-tuning the 2D Character model in \textit{Task 1}, aiming to "\textit{emphasize character clothing features}.”}
Tagging the training images can be assisted with automation tools.
However, manual refinements are necessary \revise{because appropriate captions are required to align with the user's intention, whilst current solutions fail to capture}.

\item
\textbf{Model training.}
Based on the processed data, users set hyper-parameters to start the training process.
The process will be terminated if anomalies arise (\eg, non-converging loss values), and users will adjust the dataset or hyper-parameters to restart training.
However, the training process is not transparent, as users can only monitor the progress through log data such as loss values that only indicate whether the training converges or not.
Due to the \revise{subjective} nature of \revise{the quality of image generation matching intentions}, some users choose to periodically generate sample images and break the "black box" progress.
They configure a set of intent-aligned prompts to check that current training settings are consistent with their intentions.

\item
\textbf{Model evaluation.}
Users select the best model from a series of checkpoints obtained during training.
Some checkpoints are often not satisfactory, and users need to manually adjust prompt schemes to better match user intentions based on multiple evaluation metrics.
\end{itemize}


\subsubsection{User Intention}\label{sssec:pre_intention}
\begin{figure}[t]
\centering
 \includegraphics[width=0.95\textwidth]{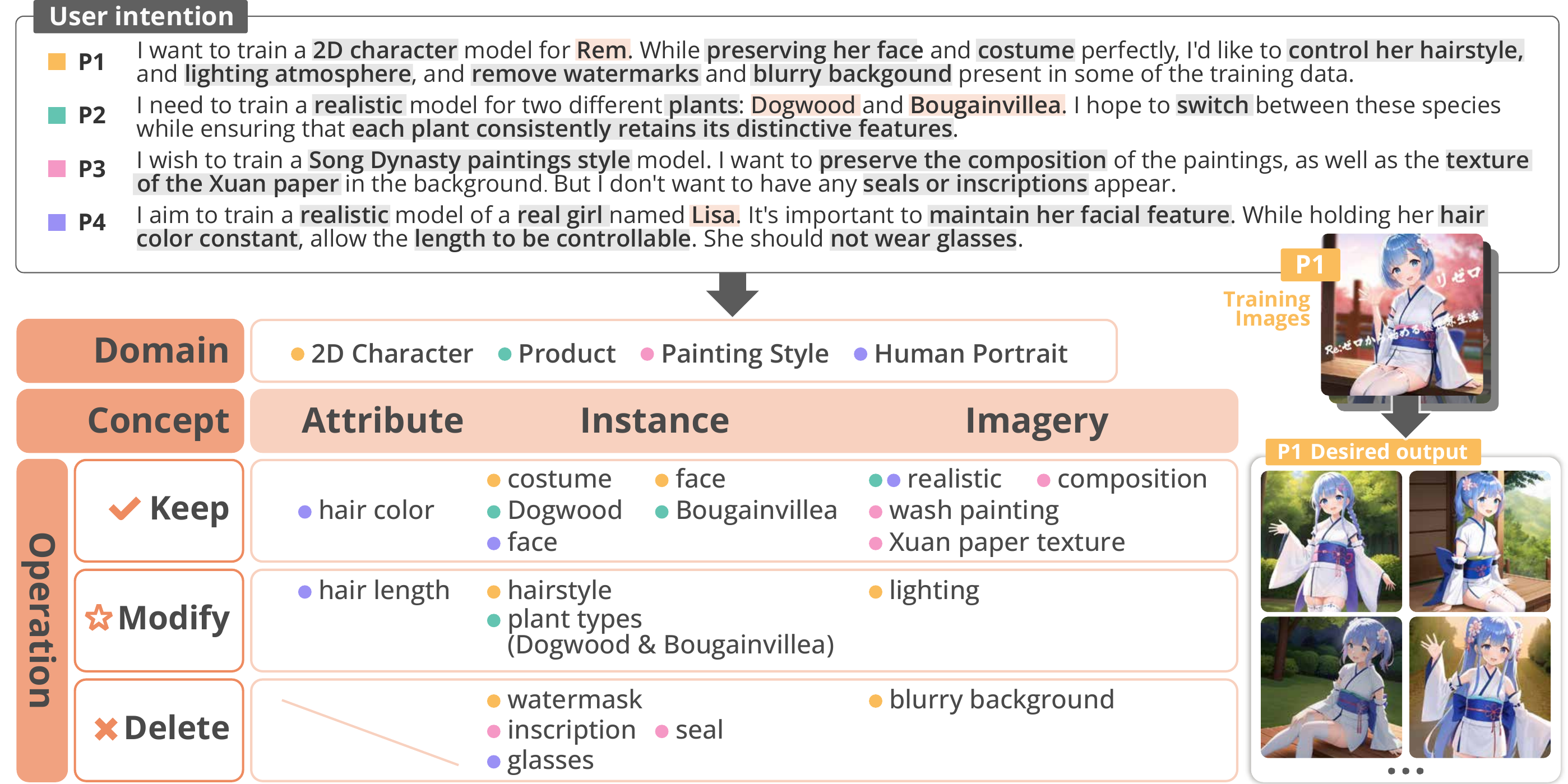}
 \vspace{0mm}
\caption{\textbf{User intention}. We summarize user intentions in \textbf{Domain}, \textbf{Concept}, and \textbf{Operation}. \textbf{Domain} refers to the specialized area of creation, such as \emph{"2D character"}. \textbf{Concept} defines the specific elements in the intentions, including three different granularities: Attribute (\eg, \emph{"hair color"}), Instance (\eg, \emph{"face"}, \emph{"costume"}) and Imagery (\eg, \emph{"lighting"}, \emph{"blurry background"}). \textbf{Operation} encompasses three different types of intended manipulations on specific concepts, including Keep, Modify, and Delete. For example, P1 wants to keep the costume, modify the hairstyle, and delete the watermarks. } 
\vspace{-4mm}
\label{fig: user intention}
\end{figure}

We summarize user intentions in \textbf{Domain}, \textbf{Concept}, and \textbf{Operation}, as depicted in Fig.~\ref{fig: user intention}.

\begin{itemize}
\item 
\textbf{Domain} designates the specialized area of creation targeted by the fine-tuning (\eg, 2D Character, human portrait, product design, and painting). 
Based on the general characteristics of the target domain, users can form an initial understanding of the training focus.
For instance, P1 stated, "\textit{The focus of a 2D character lies in its costume}," while P4 believed, "\textit{For real-human, facial features should be the primary concern}."
The pre-trained model selection and training parameters setting are affected by the domain.

\item 
\textbf{Concept} is the elements specified in the intention, such as hair length, costume, and background. 
Depending on the scope of different concepts, we classify them in ascending order of granularity: \textit{Attribute level} (\eg, hair color, facial expression), \textit{Instance level} (\eg, hair, face), and \textit{Imagery level} (\eg, background, lighting atmosphere). 
Fine-tuning usually introduces new concepts or redefines and aligns existing concepts within the pre-trained model, which are called "trigger words" by users (\eg, "Lisa," "Dogwood," and "Bougainvillea" in Fig.~\ref{fig: user intention}).


\item 
\textbf{Operation} refers to the intended manipulations of concepts in the generated results, which can be categorized into \textit{Keep}, \textit{Modify}, and \textit{Delete}.
\textit{"Keep"} signifies the desire to retain certain concepts from the training set, ensuring that they can be \revise{stably invoked by trigger words after fine-tuning.}
For instance, P1 wishes to preserve the "\textit{character's costume}," while P3 aims to retain the "\textit{aesthetic composition of Song Dynasty paintings}."
Specifically, keeping a concept at the instance level essentially means retaining all elements associated with that concept at the attribute level.
\textit{"Modify"} represents the capability to adjust certain concepts. 
Modifying a concept at the instance level can only manifest as switching to a different concept.
For example, P2 needed to "\textit{switch different types of plants}."
\textit{"Delete"} removes undesirable concepts from the training data to mitigate any negative effects. 
Only the concepts at the instance and imagery level can be deleted.
For instance, P3 tried to remove the "\textit{seals and inscriptions in painting}," and P4 preferred the character not to "\textit{wear glasses.}"
\end{itemize}

\subsubsection{Challenges in fine-tuning practice}\label{sssec:pre_challenge}

\begin{itemize}
\item 
\textbf{C1: The \revise{abstract} intentions are challenging to be translated into clear data strategies}.

Regardless of user expertise, fine-tuning often undergoes a tedious trial-and-error process.
\revise{One reason is that the training intentions are typically diverse and complex, involving varied operations for multiple levels of concepts.}
Moreover, the focus and operations in training intention vary significantly across different domains, making it challenging to adapt experience from one domain to another.
For example, P1 is "\textit{only familiar with 2D Character model fine-tuning}", and P3 specializes in "\textit{painting style}." 
As such, fine-tuning models in alignment with intentions entails substantial learning overhead and cognitive load.

\item
\textbf{C2: \revise{Insufficient} visual samples and \revise{unreliable} textual captions hinder intention alignment}.

\textit{C2.1: Insufficient visual samples}.
The quantity and quality of training images for fine-tuning tasks are usually insufficient or deficient, resulting in poor training.
For example, during a 2D Character model fine-tuning task, P1 only obtained a limited dataset of 7 images.
\revise{He had to crop the images and duplicate them} to "\textit{emphasize character clothing features}".
However, the augmentation process requires repetitive manual operations, making it tedious and time-consuming. 
Moreover, removing unintended concepts from the raw dataset can be difficult.
For example, P1 wanted to "\textit{remove watermarks and blurry background}."
He manually cropped out watermarks to "\textit{remove watermarks}," but removing "\textit{blurry background}" is challenging by editing the images manually.
Adding the concepts into negative prompts can reduce their appearance in the generated images, yet the approach only suppresses the concept's appearance without actually deleting it.

\textit{C2.2: \revise{Unreliable} auto-tagging and cumbersome manual tagging.}
Manual tagging is labor-intensive, yet \revise{auto-tagging methods indiscriminately describe image content that may produce inaccurate and non-intended tags.}
Users often experience confusion when manually adjusting the auto-tagged captions. 
They usually "\textit{do not understand the relationship between the tagging strategy and the intended result}" (P1). 
All participants found the data processing phase "\textit{tedious and challenging}" but believed that "\textit{automated methods are likely to have difficulty replacing this process due to the system's inability to understand my intentions}" (P1).

\item
\textbf{C3: Intuitive training monitoring and effective evaluation are missing}.

\textit{C3.1: Unintuitive monitoring of the black-box training process.}
All participants were concerned about the black-box training process,
They monitor anomalies through abstract parameters in the logs but can not establish an intuitive expectation of the outcome. 
Some training tools have a "sample image" feature. 
However, this feature "can only observe the effect in a preset perspective" (P3). 
Moreover, the model training process is volatile. 
P2 wishes to see "\textit{the trend over time}" to form a clear expectation.

\textit{C3.2: Lack of intention-aligned evaluation metrics.}
After training, users must evaluate the model from multiple perspectives based on training intentions to find the direction for further iterations or select a satisfactory model. 
For evaluation metrics, current practice requires users to manually adjust prompts continuously to test the model's alignment with intentions. 
For evaluation methods, participants believe that the assessment of model performance relies on human subjective feelings. 

\end{itemize}

\subsection{Design Goals}\label{ssec:pre_goal}
Based on the findings, we established three design goals to guide the development of \textit{IntentTuner}:
\begin{itemize}
\item 
\textbf{G1: Understand user intentions via natural descriptions and interactions.} The system should automatically extract intention structures from natural user input and align them with concepts in the training dataset for understanding, facilitating the translation of abstract intentions into specific system commands \revise{(C1)}.

\item  
\textbf{G2: Provide efficient intent-aligned data augmentation.}
Regarding C2.1, the system should support efficient and effective augmentation of image training samples while mitigating potential risks of overfitting.
To address C2.2, the system should incorporate intelligent caption optimizations to highlight user intentions in fine-tuning.

\item  
\textbf{G3: Offer intention-aware intuitive monitoring and evaluation of model performance. }
The system should provide intent-aware metrics (C3) and support intuitive monitoring with trend visualization of the metrics (C3.1) along with an easy-to-use generation panel for swiftly and comprehensively evaluating model checkpoints (C3.2).
\end{itemize}

\section{\textbf{Framework Overview}}
\label{Sec-framework}
Based on the identified design goals, we propose a novel framework to integrate human intent into the fine-tuning workflow.
The framework consists of three stages: 
1) understanding user intents and transforming them into structured intent specifications (Sect.~\ref{ssec:intent_specs}); 
2) enhancing training datasets with intent-guided image augmentation and caption optimization (Sect.~\ref{ssec:intent_data augmentation}); and 
3) monitoring and evaluating generated images with intent-aligned metrics (Sect.~\ref{ssec:intent_evaluation}).

\subsection{Language-Vision Intent Alignment and Transformation}
\label{ssec:intent_specs}
\begin{figure}[t]
\centering
 \includegraphics[width=0.995\textwidth]{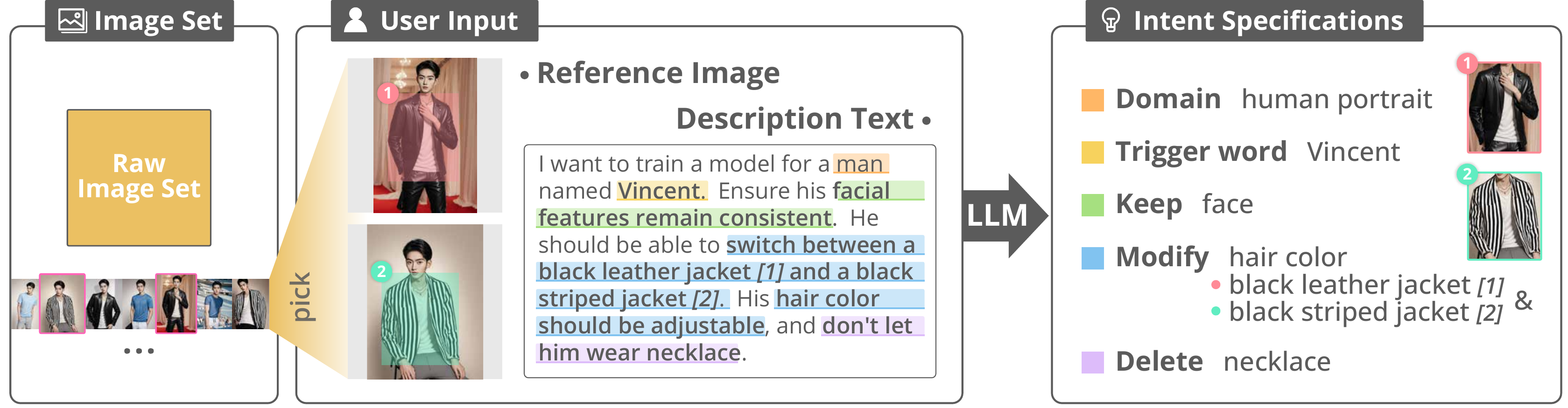}
 \vspace{0mm}
\caption{\textbf{Language-vision intent input and transformation.} We allow users to provide detailed multi-modal input to clarify their intents, including the description text and reference images. Powered by the language model, the user input will be transformed into intent specifications, including trigger words, domain, concepts, and operations.} 
\vspace{-2mm}
\label{fig:intent-input}
\end{figure}

Supporting users to express their training intent accurately with a low burden is a non-trivial task due to the multi-modal nature of fine-tuning text-to-image generation.
While natural language serves as an intuitive and widely used channel for user input, it alone cannot sufficiently describe user intents due to the complexity of training images.
This issue becomes more pronounced when some concepts in the training images share similar semantics.
For example, as shown in Fig.~\ref{fig:intent-input}, when the user intends to teach the model a new human concept while keeping his clothing, they may initially describe their intents in simple keywords (\eg, \textit{"learn the black jacket."
}).
Language-vision models can not fully clarify users' intents because there are two specific types of black jackets, namely \textit{black leather jacket} and \textit{black striped jacket}.
Conveying such detailed intentions requires more concrete textual descriptions, 
and general vision models still struggle to discriminate such fine-grained categories directly~\cite{liu2023grounding}.
Therefore, explicit correspondences between text and fine-grained visual elements are needed.
To fulfill the requirement, we construct a language-and-vision input alignment stage to assist users in articulating their intents robustly and accurately.

As shown in Fig.~\ref{fig:intent-input}, users can articulate their intentions by providing text descriptions and reference images.
Specifically, users can choose reference images from their image set and use bounding boxes to select specific visual concepts.
We set up a unique grammar to help users refer to those visual concepts in the text, indicated by numbers in brackets (\eg, "[1]"). 
For example, the user inputs 
\revise{"\textit{I want to train a model for a man named Vincent. Ensure his facial features remain consistent. He should be able to switch between a black leather jacket [1] and a black striped jacket [2]. His hair color should be adjustable, and don't let him wear a necklace.}"}
We target to extract a concrete intent hierarchy (\ie, domain-concept-operation, see details in Sect.~\ref{sssec:pre_intention}) from the multi-modal input, where operation includes \textbf{\textit{Keep}}, \textbf{\textit{Modify}}, and \textbf{\textit{Delete}}.
Among the text descriptions, "face," "hair color," "black leather jacket," "black striped jacket," and "necklace" are detected as visual concepts. 
\revise{Moreover, their associated domain (\eg, "human portrait") and trigger words (\eg, "Vincent") are detected according to the context, and image references will be attached to concepts if any number reference is provided.}
These intent specifications (\eg, Fig.~\ref{fig:intent-input} \raisebox{-.2\height}{\includegraphics[width=0.32cm]{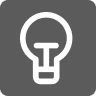}}) will guide subsequent parts of our framework.

To implement this transformation, we exploit the in-context learning capability of LLM~\cite{min2022rethinking}, which empowers the model to conduct novel tasks guided by minimal examples without specialized training for each task. 
Specifically, we enhance the robustness of intent transformation by structuring few-shot examples in the "chain-of-thought" manner~\cite{wei2022chain}, \revise{encouraging the LLM to follow a step-by-step reasoning process.} 
\revise{
In particular, chain-of-thought prompting not only clarifies the questions and requirements about the intent specifications but also provides
exemplary intermediate human-like reasoning steps (\eg, the rationale behind classifying the training domain to "human portrait") for the LLM to imitate, making complex inferences more transparent and interpretable.
}
Detailed few-shot examples for instructing the LLM are provided in the Supplementary Materials.

\subsection{Intent-guided Data Augmentation}
\label{ssec:intent_data augmentation}
Guided by the operation-concept pairs in intent specifications, we further conduct data augmentation on all raw images for intent-aligned data representations. 
Below, we introduce the image augmentation methods corresponding to different operations and present the caption generation and optimization strategies.

\begin{figure}[t]
\centering
 \includegraphics[width=0.995\textwidth]{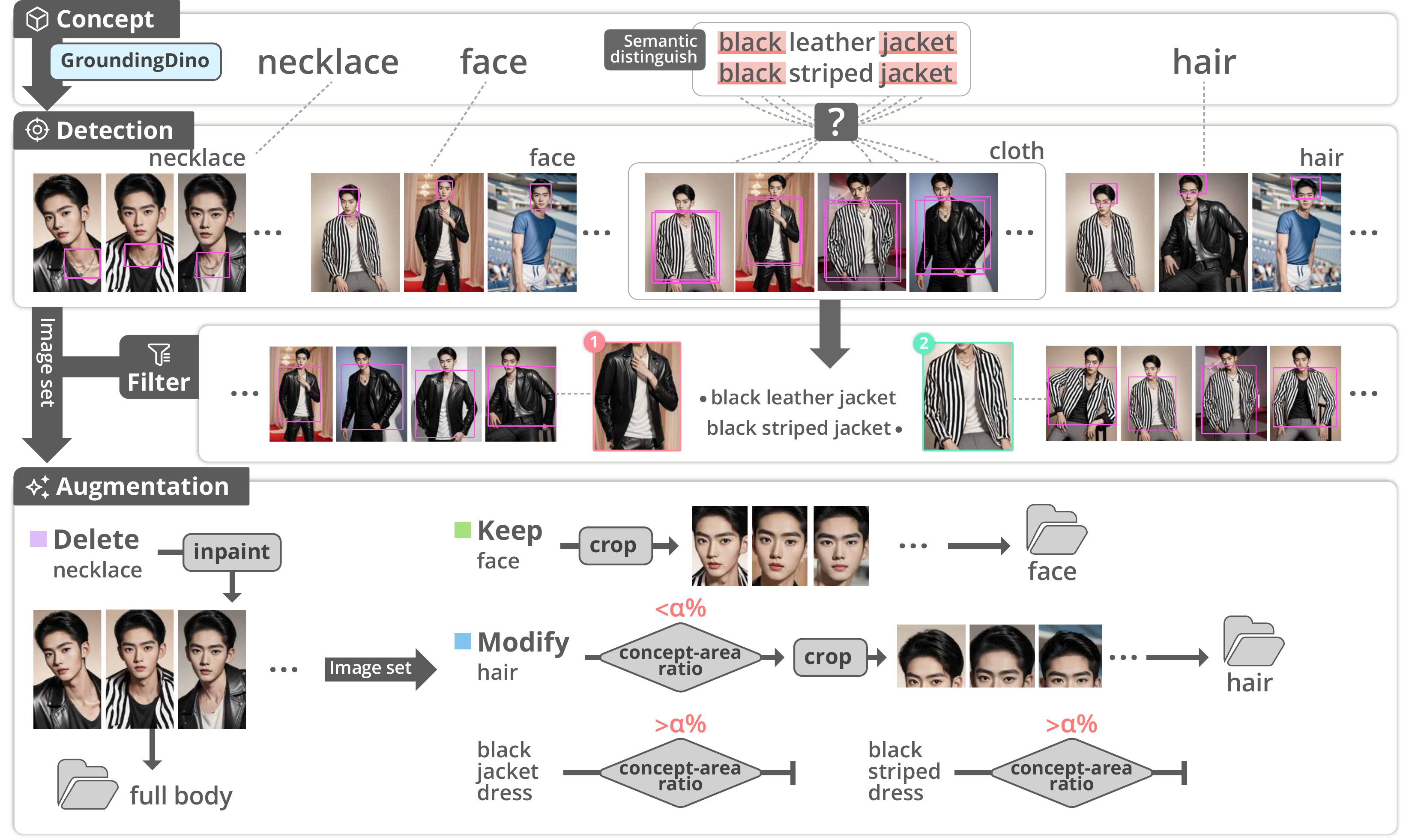}
 \vspace{0mm}
\caption{\textbf{Image augmentation.} Based on the intent specifications shown in Fig.~\ref{fig:intent-input}, we introduce a language-vision intent filter to transfer users' precise intentions to achieve intent-guided data augmentation. Specifically, the fine-grained concepts are passed to a cross-modal Detection module, which can disambiguate the intended concepts and locate the corresponding visual concepts. Then, in the Filter module, users can accurately retrieve samples with the specified concepts with the help of the reference images. Finally, the concept-aligned samples are augmented based on different intended operations to provide more intent-aligned fine-tuning data. } 
\label{fig:image-augmentation}
\vspace{-4mm}
\end{figure}

\subsubsection{Image Augmentation}
As shown in Fig.~\ref{fig:image-augmentation}, image augmentation encompasses two primary stages: 1) detection and filtering of intent-related concepts; 2) augmentation based on different operations.
Here, we first employ GroundingDino~\cite{liu2023grounding}, a pre-trained vision model that excels at identifying image regions corresponding to textual descriptions.
GroundingDino takes intent-related concepts as text input and detects the bounding boxes of their associated visual elements, as shown in Fig.~\ref{fig:image-augmentation}.
The detection results then form the basis for the subsequent augmentation process.
To address the semantic ambiguity inherent in GroundingDino's detection,  we utilize the earlier reference images as a "filter."
Specifically, we calculate the similarity between the detected concepts and the reference image and filter out the visual concepts with low similarities. 
This approach effectively mitigates the issue of incorrectly detected visual concepts that share similar semantic meanings.

Based on the detected bounding boxes, different augmentation strategies will be performed based on different intents.
\begin{itemize}
	\item
\textit{Delete} intent will require further inpainting. 
Cropping an image directly from the bounding box of a visual concept is an intuitive approach. 
However, this method \revise{may also remove visual and semantic information embedded in the cropping regions} of the original image.
To lighten such an issue, we employ inpainting techniques~\cite{rombach2022high} to remove visual concepts while redrawing the removed area based on the surrounding area, thus preserving as much of the original semantic and visual information as possible.
\item
\textit{Keep} intent will be translated as cropping the images inside the bounding boxes and adding the cropped parts as independent images in the training dataset. 
This will allow the model to pay more attention to the semantic and visual information of the concept in the following training phase.
\item
\textit{Modify} intent is similar to \textit{Keep}, which should be concerned about whether it takes an appropriate proportion of the training dataset.
Differently, we set a trigger threshold (\ie, the area percentage of the concept in the original image, default set to 40\%) for the image cropping operation based on two intuitive principles from practice. 
Firstly, too many new independent images will significantly lead to an increase in training time. 
Secondly, repeating concepts that are inherently salient in original images easily leads to overfitting.
\end{itemize}

\revise{The resulting images will constitute a new image dataset that consists of multiple sub-folders, such as \emph{face, hair}, and \emph{full body} in Fig.~\ref{fig:image-augmentation}, which will be used for subsequent training instead of using the original dataset.}
\subsubsection{Caption Optimization}
\label{ssec:caption-optimization}
Text-to-image generation is inherently a multi-modal process that requires each training image to be coupled with an intent-aligned text caption. 
Commonly used caption generation strategies leverage the CLIP series model (\eg, BLIP2~\cite{li2023blip} and CLIP interrogator~\cite{phar23clip-interrogator}) to caption images.
However, an image can correspond to vastly different captions, and such processes do not consider user intent and require further manual adjustment of generated content that significantly increases user workload.
To take advantage of the intention specifications obtained in Sect.~\ref{ssec:intent_specs}, we design intelligent intent-aligned caption optimization.
We first utilize the state-of-art captioning model BLIP2~\cite{li2023blip} to automatically generate the initial caption of the image and add the trigger word at its beginning.
Next, the caption optimization strategy will help users clarify their \emph{Keep} and \emph{Modify} operations.
\textit{Delete} operation has no caption optimization as the concept has been deleted during the image augmentation stage.

\begin{figure}[t]
\centering
 \includegraphics[width=0.8\textwidth]{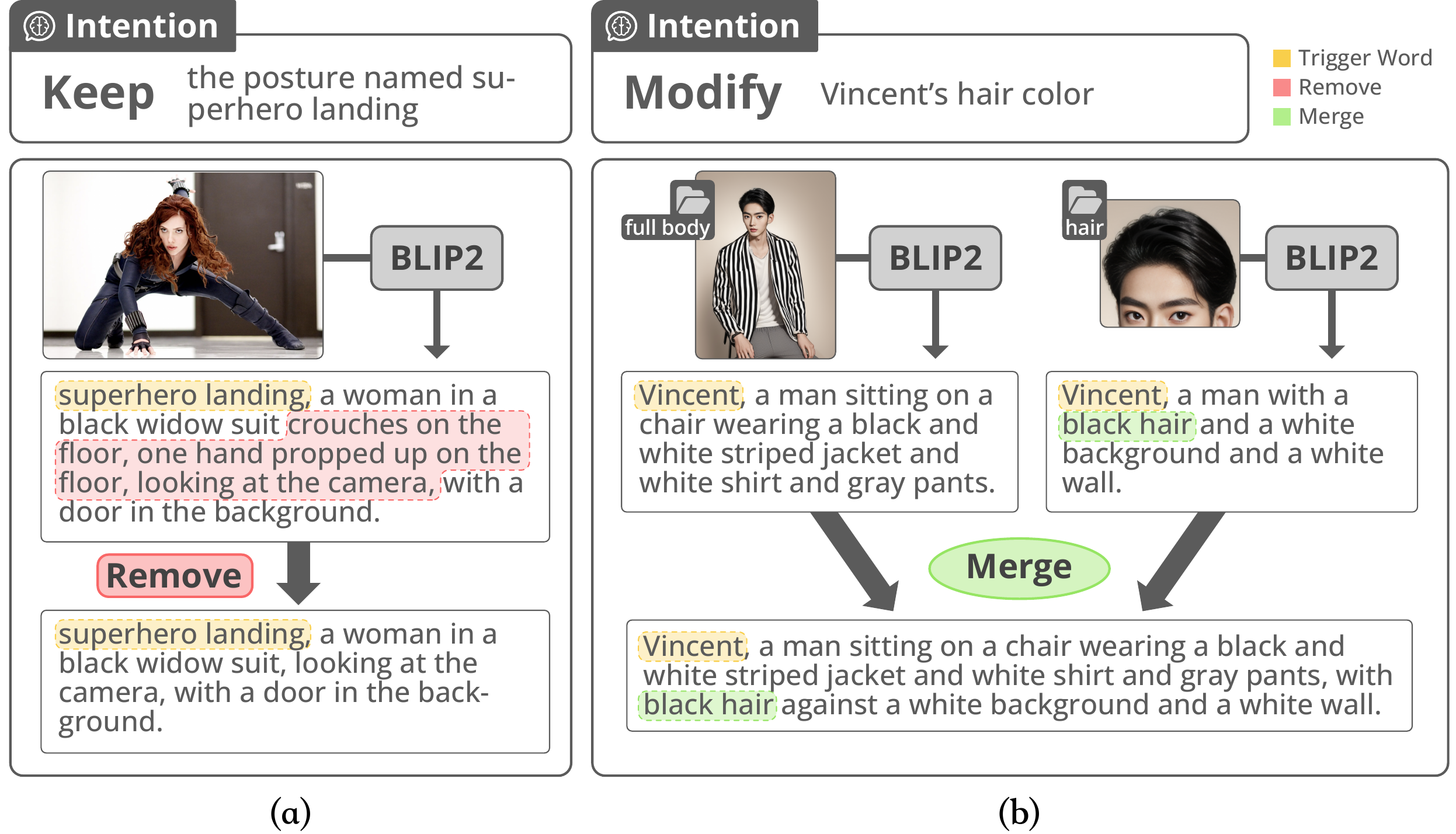}
 \vspace{-2mm}
\caption{\textbf{Caption optimization.} Caption optimization can intelligently enhance auto-generated captions based on intent specifications. (a) For \emph{Keep} intent, the optimization detects and deletes the redundant concept description to maintain an unambiguous mapping to the trigger word. (b) For \emph{Modify} intent, the optimization focuses on the target concept and generates a detailed description that complements the initial caption, to avoid the visual concept being bound to other words.} 
\vspace{-2mm}
\label{fig:caption-augmentation}
\end{figure}

\textit{Keep} means that users hope the concept can be bound to the trigger word (\emph{"superhero landing"} in Fig.~\ref{fig:caption-augmentation} (a)), so they can always generate an image containing the referred concept after fine-tuning. 
For this purpose, the caption optimization needs to identify and \revise{remove} contents in the initial caption that relate to the referred concepts so that the model can learn an explicit mapping between the trigger word and the referred visual concept.
Specifically, as in Fig.~\ref{fig:caption-augmentation} (a), users want to keep a posture called "superhero landing" and use this single phrase to prompt the fine-tuned model to generate such posture.
The initial caption automatically generated is
\textit{"superhero landing, a woman in a black widow suit crouches on the floor, one hand propped up on the floor, looking at the camera, with a door in the background"}, containing detailed descriptions of the posture (\textit{"crouches on the floor, one hand propped up on the floor, looking at the camera"}), which can confuse the generative model as to what the trigger word refers to.
Guided by the \emph{Keep} operation in the intent specifications, we prompt the LLM to locate the relevant descriptions and \revise{remove} them, yielding the optimized caption: 
\textit{"superhero landing, a woman in a black widow suit, looking at the camera, with a door in the background"}. 
As a result, the special pose will be bound to the trigger word and not be confused with other descriptions.

\textit{Modify} is incorporated to control and change certain concepts in the fine-tuned model.
For this purpose, the caption should include a detailed description of the concept to avoid its visual features being bound to the trigger word or other parts of the caption.
However, in some cases, the generated caption may overlook the concept and not provide a detailed description.
For example, as shown in Fig.~\ref{fig:caption-augmentation} (b), users want to control the portrait's hair color.
However, the automatically generated caption completely overlooked the visual concept.
To describe the hair attribute in detail, we leverage the previous concept detection results to focus on the hair and prompt BLIP2 to generate a detailed caption.
Next, we prompt the LLM to merge the initial caption with the detailed caption to produce the final optimized caption, successfully incorporating the user's intent into the captions.

\subsection{Intent-aligned Image Evaluation}\label{ssec:intent_evaluation}
We aim to allow users to assess the alignment between model output and their intents quantitatively rather than merely observing images subjectively. 
Specifically, we evaluate the effect of text-to-image fine-tuning by measuring the generated images from two complementary aspects: stability and controllability.
\revise{
Stability measures the model's ability to replicate the target concept~\cite{gal2022image}, \ie, whether the target concepts in the generated images are visually similar to users' expectations.
However, only considering stability may cause overfitting, as the model may only remember and repeat the training data in extreme cases and lose diversity in other visual properties~\cite{somepalli2023diffusion}.
For instance, for an overfitted model, regardless of how users set the positive and negative prompts, users cannot obtain specific properties (\eg, "\emph{long hair}"), and all the generated images end up being similar, such as having short hair.
In other words, users lose control of other properties like hair via prompts because the overfitting leads to mode collapse.
To mitigate this issue, we further develop controllability to measure the model's ability to modify the concepts using textual prompts.
For instance, hair length controllability refers to whether users can manipulate the output images to have varying hair lengths by changing the prompts. 
}

\noindent \textbf{Stability.}
\revise{We evaluate model stability at the granularity of user-desired concepts, not limited to the overall image. 
This allows users to evaluate multiple intents independently.
}
Specifically, we crop intent-related objects and compute the visual similarity between them and each generated image.
Conventional methods that transform images into high-dimensional vectors and compute vector similarity based on a pre-trained feature extractor (\eg, CLIP-Vit and CLIP-ResNet~\cite{radford2021learning}) neglect the perception of human preferences~\cite{xu2023imagereward}.
Inspired by~\cite{wu2023better}, we leverage the human-preference classifier to mitigate the issue, which is fine-tuned on a large-scale human-labeled dataset that focuses on discriminating the common drawbacks of the generated images compared to the real ones.
Specifically, 
\begin{equation}
\mathrm{Stability} = \frac{1}{NM} \sum_{i=1}^{N} \sum_{j=1}^{M} \mathrm{sim}(I_i, R_j),
\end{equation}
where $I$ refers to the sampling image set during training and  $M$ is its batch size.
$R$ refers to the intent-related object set, $N$ is its total number, while $sim(\cdot)$ refers to the similarity calculation based on the human preference classifier~\cite{wu2023better}.

\noindent \textbf{Controllability.}
To evaluate whether we can modify specific concepts as expected, we first use prompts containing opposite attributes of the specified concept (\eg, short hair \emph{vs.} long hair) to generate sampling images.
Then, we measure the standard semantic alignment metric in a high-dimensional, non-linear neural embedding space like CLIP to represent text-image similarity. 
Inspired by previous research~\cite{feng2023promptmagician}, we leverage the two opposing attribute keywords to transform the image evaluation task to a binary classification, which can effectively reduce the ambiguity that arises from using a single attribute~\cite{wang2023exploring}.
We then calculate the latent space cosine similarity of each image with the two opposing text keywords of user-intended control attributes ($sim_1$ and $sim_2$, where $sim_1$ denotes the similarity to the correct control keyword and $sim_2$ denotes the similarity to the opposing keyword). Next, we compute the controllability score as the normalized Softmax similarity toward the intended attributes:
\begin{equation}
\mathrm{Controllability} = \frac{1}{M}  \sum_{i=1}^{M}  \frac{e_i^{sim_1}}{e_i^{sim_1}+e_i^{sim_2}} , 
\end{equation}
where $M$ is the batch size of sampling images, and $sim$ refers to the text-image similarity.

\section{System Design}
We develop an interactive system to support user-intended fine-tuning of text-to-image models.
This section introduces the system's interface and interaction design to accomplish: 1) user intent understanding (Sect.~\ref{ssec:understanding-user-intent}), 2) automated data preparation (Sect.~\ref{ssec:intent-guided-data-strategy}), and 3) intent-aligned monitoring and evaluation (Sect.~\ref{ssec:system-evaluation}), as described above. 
We associate them with the design goals in Sect.~\ref{ssec:pre_goal} to underscore the design rationale.

\begin{figure}[t]
  \includegraphics[width=0.99\textwidth]{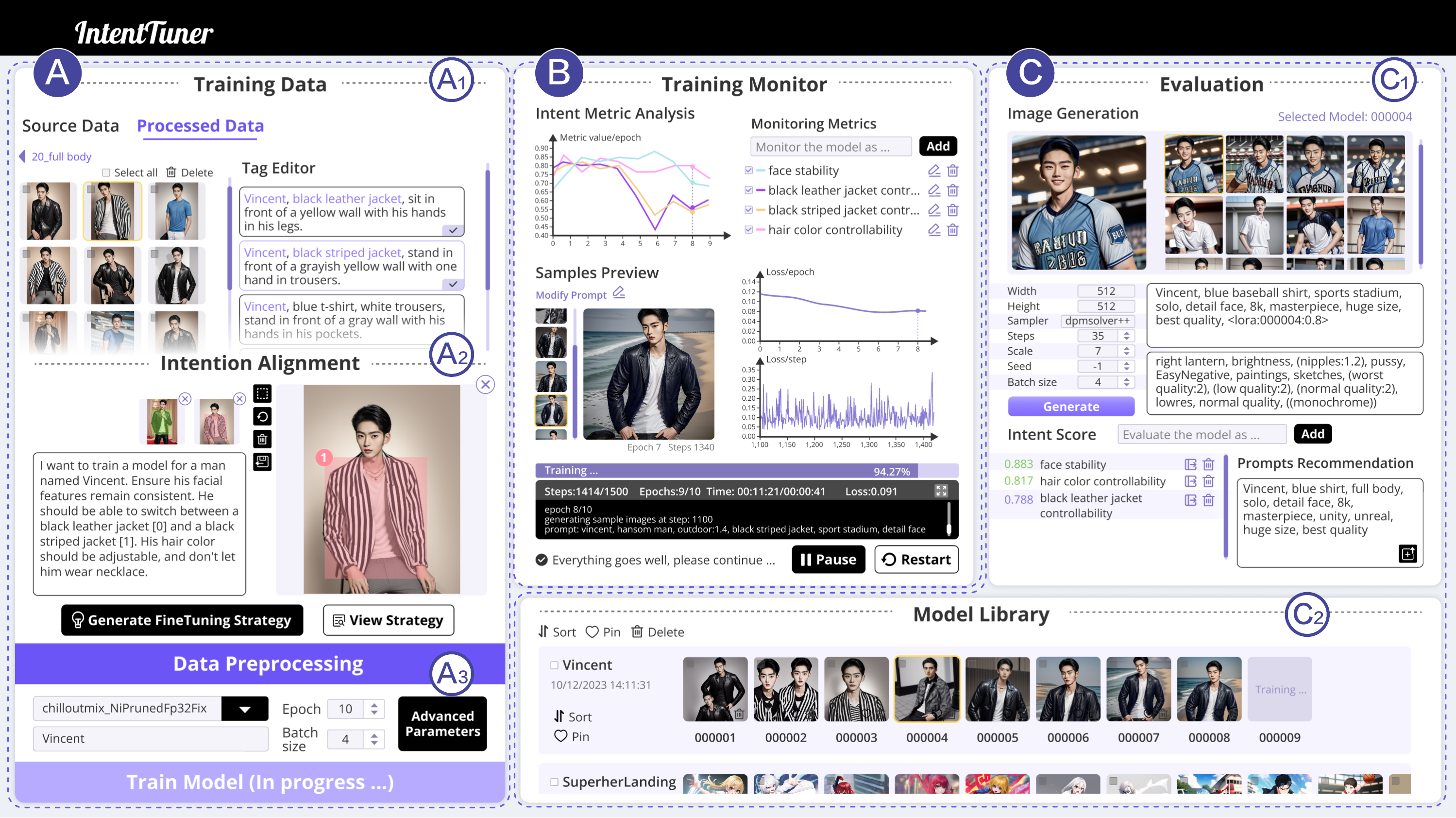}
  \caption{\textbf{User interface of \textit{IntentTuner}.} The \textit{Intention-Data Alignment Module} (A) allows the user to input the model's fine-tuning training data and intentions, conduct pre-processing, and configure other training settings and hyperparameters. The \textit{Training Monitor Module} (B) monitors and visualizes training progress based on intentions. The \textit{Model Evaluation Module} (C) helps users evaluate models based on multiple metrics from intentions.}
  \vspace{-4mm}
  \label{fig: interface}
\end{figure}

\subsection{User Intent Understanding}
\label{ssec:understanding-user-intent}
The \textit{Intention Alignment} panel (Fig.~\ref{fig: interface} A2) allows users to input their intentions naturally, assisting them in materializing the abstract training intention (G1). 
First, users upload their training image set in the Source Data page in the \textit{Training Data} panel (Fig.~\ref{fig: interface} A1). 
Subsequently, they can describe training intentions using natural language in the text area of the \textit{Intention Alignment} panel (Fig.~\ref{fig: interface} A2).
After that, to accurately refer to some visual concepts in the text description, users can drag images containing relevant concepts from the training set into the image canvas. 
The image canvas offers a selection tool 
\raisebox{-.2\height}{\includegraphics[width=0.32cm]{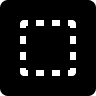}}
to highlight concepts, and it will automatically number the user's selections, distinguishing them with different colors.
Other functions for canvas annotation are provided, such as \textit{undo}~\raisebox{-.2\height}{\includegraphics[width=0.32cm]{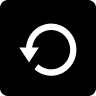}}, \textit{clear all}~\raisebox{-.2\height}{\includegraphics[width=0.32cm]{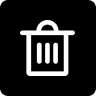}}, and \textit{save}~\raisebox{-.2\height}{\includegraphics[width=0.32cm]{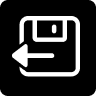}}.
Users can drag in multiple images for annotation, facilitating the training task with multiple concepts. 
Then, users can reference the annotated concept in the textual description using square brackets "[]", enabling references between text and specific visual concepts.
Finally, users can click the  \textit{Generate FineTuning Strategy} button~\raisebox{-.2\height}{\includegraphics[width=0.32cm]{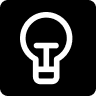}} to automatically translate the intention input into intent specifications (\eg, domain, trigger word, and operation-concept pairs) and recommend prompts needed in subsequent image generation testing accordingly. 
\revise{The intent specifications can be viewed in a JSON format via the "View Strategy" button~\raisebox{-.2\height}{\includegraphics[width=0.32cm]{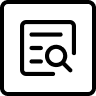}}, and users can edit them flexibly to increase the controllability of the system's behavior and enhance user trust.}

\subsection{Automated Data Preparation}
\label{ssec:intent-guided-data-strategy}
When the user's intention aligns with the system's understanding, they can preprocess the entire dataset through the \textit{Data Preprocessing} button. 
Guided by the intent specifications, the system will automatically perform cropping, inpainting, and classification of the image set and image captioning (G2).
The newly processed dataset will be displayed on the Processed Data page in the \textit{Training Data} panel (Fig.~\ref{fig: interface} A1). 
Users can view the system's image processing results in this panel, including classified folders and cropped images, and modify the captions. 
However, the amount of captions in a training set is enormous. 
To save users' workload in caption modification, the \textit{Tag Editor} supports the flexible propagation of modifications to a specific caption to other captions. 
In addition, content related to intentions will be highlighted in the captions, helping users efficiently locate key focus information and enhancing their understanding of the relationship between captions and intentions. 
\revise{The design principle behind the \textit{Tag Editor} is providing appropriate user feedback, ensuring the system fully understands user intent before fine-tuning.}

\revise{
The \textit{Settings} panel (Fig.~\ref{fig: interface} A3) displays the necessary settings for model training (\eg, base pre-trained model and trigger word). 
The system presets training hyperparameters based on the fine-tuning domain identified from user intents.
The system hides hyperparameters other than batch size and epoch in the Advanced Parameters button, which users can click to view and adjust more detailed parameters.
For more details and illustrations about default hyperparameters (\eg, optimizer, scheduler), please refer to Appendix~\ref{appendix-lora}, as different domains correspond to different settings.
}

\subsection{Intent-aligned Monitoring and Evaluation}
\label{ssec:system-evaluation}
After preparing the training data, users can click the \textit{Train Model} button to start training. 
Subsequently, the \textit{Training Monitor} module (Fig.~\ref{fig: interface} B) is activated. 
Specifically, the \textit{Intent Metric Analysis} panel visualizes monitoring metrics based on intentions. 
These metrics are automatically generated based on the system's understanding of intentions, which users can also continue to add or modify. 
The trend of all metrics is displayed on the same coordinate using line graphs of different colors, helping users form overall expectations of the progressive and fluctuating progress of model training. 
In addition, the \textit{Samples Preview} panel continuously displays generated images more intuitively to show the training progress. 
The initial image prompt is automatically generated based on intentions, but users can also manually change the prompt at any time for more flexible observation.

Meanwhile, the checkpoint models generated during training are stored in real-time in the \textit{Model Library} (Fig.~\ref{fig: interface} C2), and the sample image corresponding to the model is automatically set as its cover to help users browse. 
Users can select models from the \textit{Model Library} and perform image generation testing in the \textit{Evaluation} panel (Fig.~\ref{fig: interface} C).
The \textit{Intent Score} panel automatically provides intention-related metrics and their performance scores. 
Users can also continue to add, delete, or modify metrics. 
All metrics are sorted from high to low based on their scores, making it easy for users to learn about the strengths and weaknesses of the current model. 
Based on the metric content, the system can recommend prompts for image generation, and then models can be tested through the \textit{Image Generation} panel (Fig.~ \ref{fig: interface} C1).

\section{Evaluation}
The key innovations of \textit{IntentTuner} fall into two categories: 1) a fine-tuning framework that intelligently transforms user intentions into intent-aligned data strategies and 2) an integrated system that unifies fine-tuning and generation, supporting both expert and novice users to customize text-to-image generation models flexibly.

To verify them, we conducted evaluations through two studies, respectively.
In \textbf{Study 1}, we evaluated the framework across two general fine-tuning application scenarios: 1) abstract concept preserving and 2) multiple concepts augment and modification.
We compared the outcomes our intent-aligned fine-tuning technique produced for each scenario with those from baseline systems.
In \textbf{Study 2}, we conducted a user study, engaging in comparative tasks with a widely-used fine-tuning baseline pipeline which requires users to combine Koyhass~\cite{kohyass2022} and Stable Diffusion Web UI~\cite{sdui2022}. 
By analyzing participant questionnaires and interview feedback, we measure the functionality, overall effectiveness, and usability of the system and baseline, affirming its enhancement to the user experience.

\subsection{Application Examples}
\begin{figure}[t]
\centering
 \includegraphics[width=0.85\textwidth]{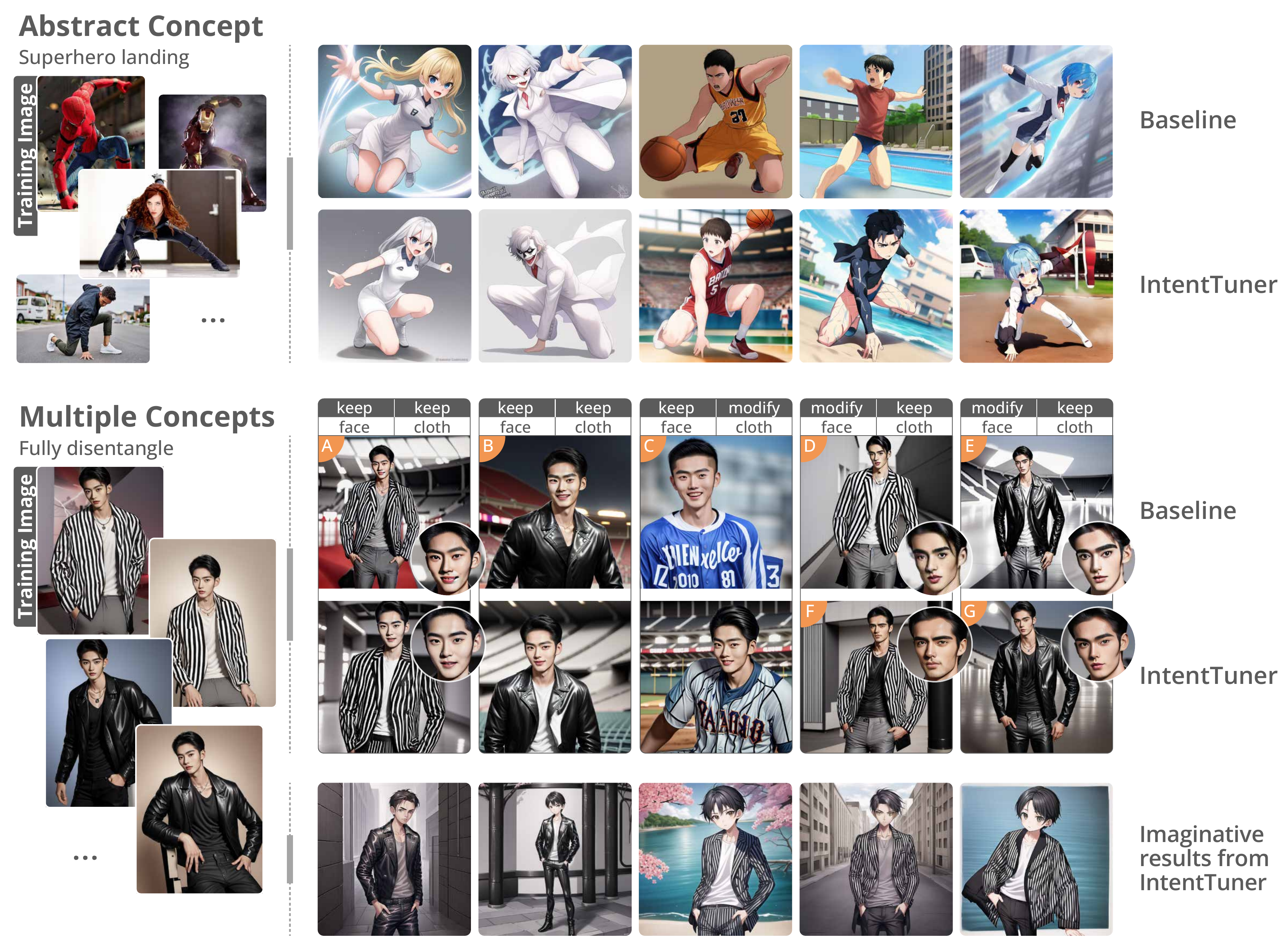}
 \vspace{0mm}
\caption{\textbf{Two general fine-tuning usage scenarios.} In each case, the first line presents results from our system, while the second line presents results without an intent-image alignment module. Interesting application results are added in the third line for the multiple concept scenario. } 
\label{fig: application-examples}
\vspace{-4mm}
\end{figure}

\noindent \textbf{Abstract concept.} 
As shown in Fig~\ref{fig: application-examples} (top), the original training dataset includes several portraits with a common pose, namely "superhero landing." 
Users intend to teach the model such an abstract concept, enabling it to
synthesize images of different people in that pose in interesting scenarios
We observe that the baseline resulted in images that can learn the "landing" concept, but could not accurately mimic the essence of the pose, \ie, placing one hand on the ground and positioning legs in front and back.
In contrast, our approach can apply it stably to objects through our intent-aligned caption optimization strategy.
The reason behind the success is to delete all information related to the human pose in the captions (see Sec.~\ref{ssec:caption-optimization} for details).
Other examples of abstract style include painting styles, photo filters, \etc, which often share the same optimization strategy and training parameters. 

\noindent \textbf{Multiple concepts.}
Fig~\ref{fig: application-examples} (bottom) shows a complex but, in practice, a rigidly necessary scenario that requires the model to learn several key concepts from a diverse dataset and to distinguish between them.
Practical applications, such as e-shopping models and product graphic design~\cite{liu2023application}, often fall into this category.
Here, we present a complex example with the intent of
"\textit{To learn the model's looks, as well as the product features of \revise{the black leather jacket and the black striped jacket}, and to be able to support switching between and combining these new concepts without being uncontrollably bound}."
That means the user can prompt to generate images that contain these originally bound concepts, \ie, (portrait face, black leather jacket) and ( portrait face, black striped jacket), independently and rebind them on demand, which is also referred to as "disentanglement"~\cite{zhou2023clip} in deep learning.

\revise{In the results, first, we observe that our intent-aware data augmentation can improve the training quality of individual concepts.
For example, in Fig.~\ref{fig: application-examples} (A) and (B), where the user tries to generate the original portrait and clothes with baseline, the quality of the face is worse than ours.
Second, the problem of conceptual entanglement between the original portrait's face and clothes can be alleviated with our tool. 
When users intend to modify one of the concepts, such as changing the face or the clothes while keeping the other concept, the features of the original portrait will still affect the generation, directly leading to the collapse of the other portrait.
For example, as shown in Fig.~\ref{fig: application-examples} (C), when the user attempts to replicate the face but to change the clothes to a sports shirt, the face in the baseline also becomes less similar to the desired original portrait compared to our result.
Moreover, we can observe that the facial features of the portraits in the baseline crumble significantly when the user tries to make the other portraits wear target clothes, as shown in Fig.~\ref{fig: application-examples} (D) and (E) in the baseline.
In contrast, with our tool, the two costumes can be more effectively combined with other portraits, including completely different live-action portraits, as shown in Fig.~\ref{fig: application-examples} (F) and (G) where the user successfully controls the face to look like "European."
}
This case demonstrates the strong controllability of the fine-tuned model, which benefits from \textit{IntentTuner}'s powerful data augmentation strategies.
In addition, the fine-tuned model can generate learned clothes and synthesize them with imaginative concepts (see the third row of Fig.~\ref{fig: application-examples} (bottom)), thus taking full advantage of the generative model.

\subsection{User Study}
We designed a qualitative study and recruited participants to assess their user experiences and evaluate the quality of the fine-tuned model results compared to the text-to-image model fine-tuning tool widely adopted within the model community.
Specifically, we aim to evaluate and compare with the baseline system 1) the usability and usefulness of the intention-aware fine-tuning framework and 2) the overall system's support for fine-tuning practices.

\subsubsection{Participants}
We recruited 12 post-graduate students (5 females, 7 males, Average age 23.0) through recruitment messages posted on social media platforms. 
The participants are from various disciplines, including art, design, and engineering.
To showcase the versatility of our framework among various user levels, we enrolled six experts, including three accustomed to using training interfaces, three AI developers familiar with code-level operations, and six novices with no fine-tuning experience but willing to fine-tune in the future. 
All participants have used the text-to-image generative model.

\subsubsection{Baseline System}
To establish a baseline for comparison against \textit{IntentTuner}, because there is no integrated system that allows users to perform fine-tuning and generation at one stop, we combine two systems widely used in the AIGC community: Kohya-ss~\cite{kohyass2022} for the training phase and Stable Diffusion Web UI~\cite{sdui2022} for the evaluation phase of our system.

Kohya-ss is a popular open-source project designed for Stable Diffusion trainers and has received 5.7k stars on GitHub. 
It provides an interface for model training, which comprises panels of pre-trained model selection, path setting, and training parameters.
Stable Diffusion Web UI is an open-source user interface for image generation and has received 102k stars on GitHub. 
It features a suite of image generation functionalities, including Text to Image, Image to Image, and InPainting, \etc. 
In this study, we leverage its Text-to-Image feature, which encompasses modules for prompt input, parameter setting, and image generation.
To ensure a fair comparison, both \textit{IntentTuner} and the baseline were set up using the same dataset and pre-trained model, and they carried out training tasks within the same domain.

\subsubsection{Procedure and Task}
\begin{itemize}
\item 
\textbf{Introduction.}
We first provided a brief introduction of the research background. 
Next, we gathered the demographic information from the participants and asked for their consent to record their operations and results for further analysis. 
We then introduced the interface of both our system and the baseline system and demonstrated their usage through a mock dataset.
If expert users were already acquainted with the baseline system, they could ask to skip its introduction. 
Then, we presented the training dataset required for the training tasks to the participants and explained the training objective. 

\item 
\textbf{Task Design. }
To delve deeper into the evaluation of \textit{IntentTuner}'s performance, we designed a model fine-tuning task for users that combines various intentions, containing multiple requirements from the general application scenarios mentioned in Section 6.2.

We provided participants with the following training requirements: 
\textit{Please train a model for the real-life girl Sophia. 
Ensure that her facial features remain consistent and allow for hairstyle adjustments, but she shouldn't wear a hat. 
Additionally, Sophia has two outfits, including Outfit 1 and Outfit 2, and the model should be able to switch between them.}

They were asked to organize the training intention based on the given requirements and complete the fine-tuning task using the \textit{IntentTuner} and baseline systems, respectively. 
They followed the think-aloud protocol during tasks and were free to ask questions at any point. 
To measure the task efficiency of users with different expertise levels across the two systems and to ensure sufficient exploration for participants to allow them to assess each system in terms of required cognitive effort subsequently, we did not set a time limit.

\item 
\textbf{Questionnaire and Interview.}
Upon finishing the task, each participant completed a 7-point Likert scale questionnaire regarding the system's functionalities and overall performance.
In the system features evaluation, we discussed from the perspectives of intention understanding, data augmentation, and intuitive metrics. 
For the overall system usability evaluation, we asked for ratings on four aspects: easiness of use, usefulness, flexibility, and engagement.
While answering the questionnaire, participants were asked to continue with the "think-aloud" protocol to explain the rationale behind their ratings. 
Subsequently, a semi-structured interview was conducted, discussing their satisfaction with the training results, their impressions of the system, and expectations and suggestions to further gauge the user experience with \textit{IntentTuner}.
On average, each participant spent 70 minutes on the entire study.

\end{itemize}

\subsubsection{Results}

All participants completed both tasks, the following questionnaire and interview. 
We first report the participants' responses to the functionalities of the system and our observations on their fine-tuning practices. 
We then discuss the system's overall usability and the limitations we learned from the participants.

\begin{figure}[t]
\centering
 \includegraphics[width=0.995\textwidth]{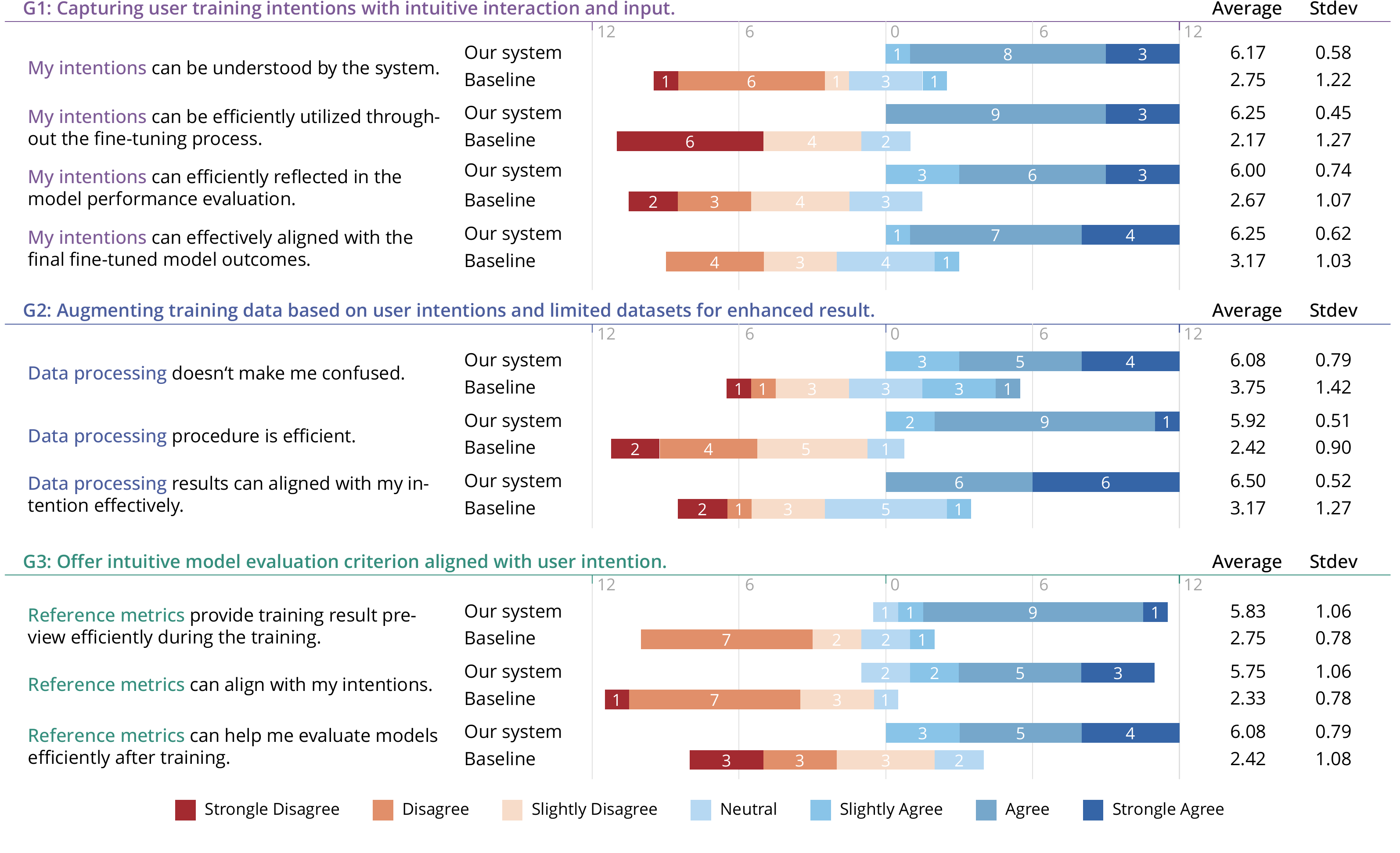}
 \vspace{0mm}
\caption{\textbf{The results of the questionnaire regarding the features experience of our system and baseline system.} } 
\label{fig: user study result_features}
\end{figure}

\begin{figure}[t]
\centering
 \includegraphics[width=0.995\textwidth]{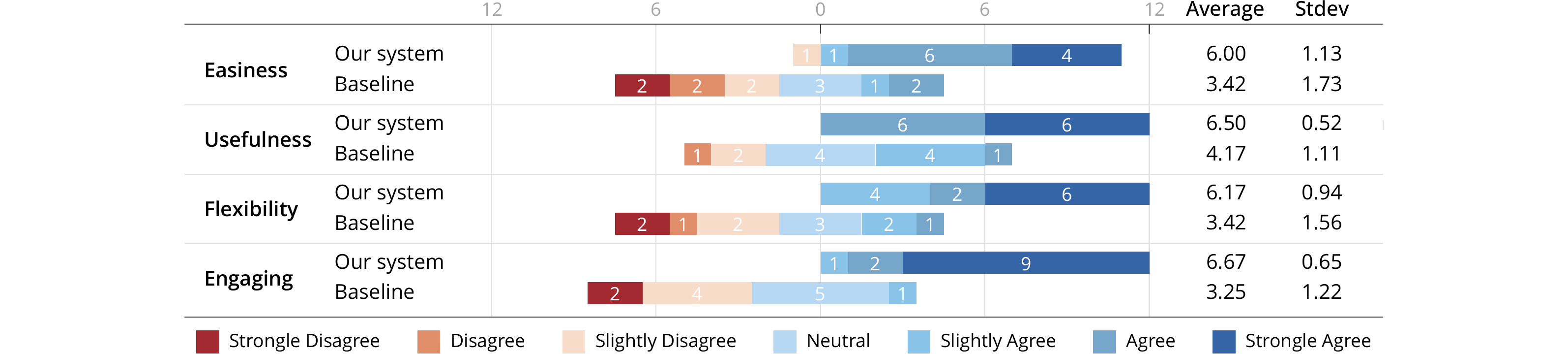}
 \vspace{0mm}
\caption{\textbf{The results of the questionnaire regarding the system impressions of our system and baseline system.} } 
\label{fig:user study result_system}
\end{figure}

\begin{itemize}
\item 
\textbf{Feedback on User Intention Capture.}
Participants found that our system can effectively comprehend their training intentions and manifest throughout the fine-tuning process, enhancing the final model quality aligned with intention (Fig~\ref{fig: user study result_features}, top).
Using natural language and reference images to express intentions is perceived as an intuitive and natural way of communication. 
This aligns with our first design goal (G1), which is to allow users to input their intentions effortlessly. 
This approach "directly reflects the high-level intentions during fine-tuning" (P1), especially when users are fine-tuning models for creative and hobbyist purposes, enabling them to "feel more free without being overly concerned about its feasibility" (P3).

Participants also noted that the system's accurate understanding of their input intentions contributed to building their confidence in the current fine-tuning task. 
P1, during her initial attempt to input intentions, had some confusion due to uncertainties about the system's granularity in understanding natural language inputs. 
However, after reviewing the examples we provided and giving it a try, she felt the intention input was "flexible and robust." 
After inputting their intentions, participants would use the View Strategy feature to assess how their intentions were understood. 
This feature offered them a "window to ensure the correct comprehension of intentions," reinforcing their understanding and trust in the system's behavior.
Participants thought the system was "simpler and more intuitive" (P3) in fine-tuning text-to-image models than the previous workflow.

\item 
\textbf{Feedback on Data Augmentation.}
Participants believed that our automatic data augmentation could easily, efficiently, and accurately integrate the embedded user intentions into the training dataset (Fig~\ref{fig: user study result_features}, middle).
In comparison, even the AI expert users found the baseline system's data processing inefficient. 
They believed that "manually processed results might not fully align with training intentions" (P7).
The intent-aligned data augmentation enabled them to "execute intentions more effectively," addressing our second design objective (G2). 
The automated process marked a significant improvement over the generic data processing workflow conducted with the baseline system. 
Participants estimated that manual data processing using generic workflows would require substantial mental and manual effort, "possibly taking 1.5-2 hours" (P1). 
For novice users, they "might not know how to align intentions with data processing methods without enough training" (P8).

After reviewing our automated data processing results, expert users believed the outcomes to be "accurate and comprehensive" (P10). 
P9 stated, "I now feel confident about the automated data augmentation." 
Additionally, they deemed the caption modification feature as "essential," and the automatic propagation of the modified caption content "saved a lot of tedious steps" (P6). 
For novice users, they didn't delve much into the data-processed results but focused more on evaluating the outcomes of the training results. 
After comparing the training results between the baseline system and ours, all participants agreed that our training outcomes "better reflected the training intentions" (P7).

\item 
\textbf{Feedback on Metrics Reference.}
Multiple participants (7/12) mentioned that our system provided intention-aligned reference metrics during both the training process and the evaluation of training results (Fig~\ref{fig: user study result_features}, bottom), addressing our third design goal (G3).

Participants believed that our reference metrics could visualize the training progress trend in a user-friendly manner. 
"The visualization of multiple intent-aligned metrics in the same coordinate provides an intriguing reference, allowing me to easily estimate the model's overall situation" (P7). 
P7 further mentioned, "Based on the 'pink dress' prompt, I could observe the color transitioning from a vibrant pink to a hue closer to that in the training set, giving me a general understanding of the training progress." 
The Model Library and Evaluation panel also integrated the model training and image generation processes. 
P5 felt that "not having to move models and switch tools manually is very convenient." 
Participants enjoyed testing images on the fine-tuned models after some training time, especially novice users who haven't yet experienced selecting models based on training epochs and other parameters. 
"This makes full use of the waiting time during training," (P1) noted. 
Specifically, P1 pointed out that the feature to view image effects during training "is not only suitable for text-to-image model fine-tuning but also applicable to all AI models that target image generation, such as image-to-image.”


\item 
\textbf{Experience of Overall System.}
Overall, participants appreciate the usability of our system more than the baseline system.
In terms of the easiness of use, our system ($mean=6.00, std=1.13$) scores higher than the baseline ($mean=3.42, std=1.73$).
As P9 commented, \emph{"even though I’m familiar with \revise{prototype AI model applications built using Gradio, I think the Gradio style is more suitable for demonstrating simple model features with fewer control settings.} But in the fine-tuning tasks, it shows too many setting parameters, requiring me to scroll up and down the interface to set different parameters, adding to my burden."}
Concerning the usefulness, our system ($mean=6.50, std=0.52$) also outperforms the baseline ($mean=4.17, std=1.11$).
Regarding the flexibility, our system ($mean=6.17, std=0.94$) is also better than baseline ($mean=3.42, std=1.56$).
\emph{"The IntentTuner system enables me to flexibly adjust my fine-tuning in more diverse and intuitive textual and visual interactions, while in the other system, I can only provide ready-made training data and adjust the caption because I don't understand the other complex settings,"} P6 remarked.
Furthermore, participants also rate our system ($mean=6.67, std=0.65$) as more engaging than the baseline ($mean=3.25, std=1.22$).
P10 said, \emph{"IntentTuner system allows me to intuitively see samples and how much they match my fine-tuning goal during the training, making the process more engaging."}
Moreover, P5 noted, \emph{"In IntentTuner, I can directly see the model gallery and instantly use different fine-tuned models to try out the generation, making the experience much more engaging than the other system, which requires me to manually select and load individual model in another UI to test the generation."}

\item 
\textbf{System Limitations}.
Even though \textit{IntentTuner} can support most fine-tuning requirements, when the user intentions become too complex, such as involving "multiple operations on multiple concepts" (P2), the training time and the uncertainty of results will increase, which is a common problem for existing fine-tuning practice.
In addition, some participants (3/12) think we should further improve the interactions in the system to increase the sense of control.
For example, P9 noted, \emph{"After I clicked the training button, I did not have much to do except wait for the results and see some intermediate samples. The monitoring is good, but I think you could try to improve this process by adding more interactions that can be performed in parallel."}

\end{itemize}

\subsubsection{Summary}
All participants were excited about the capacity of \textit{IntentTuner} to improve their workflows when fine-tuning text-to-image models. 
They agreed that \textit{IntentTuner} enables them to quickly realize their training intentions, achieve complex composite training objectives, and span various fine-tuning domains.
Specifically, the intention-oriented fine-tuning framework enables participants to be objective-driven, achieving rapid and flexible fine-tuning of text-to-image models. 
The study also revealed that \textit{IntentTuner} still has limitations in dealing with highly complex intentions and providing more user interactions during the training. 
\section{Discussion}
\revise{
\subsection{Customized AIGC: Ethical Risks and Responsive Strategies}
While the benefits of advanced generative capabilities are undeniable, their impressive power, coupled with enhanced customization options, also introduces significant ethical concerns, particularly concerning misinformation and intellectual property rights.
For instance, the ease of fine-tuning models could lead to a surge in deepfake creations~\cite{westerlund2019emergence}, enabling users to generate unauthorized synthetic representations of real individuals, especially celebrities. 
These representations risk being misused to disseminate false information, potentially causing harm to those depicted. 
Additionally, the artistic community raises alarms over copyright infringement, as their creations are increasingly harvested without consent for model fine-tuning purposes~\cite{lan2023innovation}.
These issues may also, in the long term, exacerbate biases in AI-generated content, such as the gender discrimination caused by excessive use of female figures and biased art generation due to the abuse of popular contemporary artists' styles, limiting the diversity of creation.

These ethical risks require more transparent and regulated data usage in the AIGC era. 
On one hand, legislation should keep up with the rapid development of generative AI.
Particularly, the legal definition of what constitutes a misuse of personal or intellectual property data needs to adapt to AI's increasing capacity for learning and generation.
On the other hand, advanced data security technology needs to be developed as a countermeasure.
In this regard, both preventive and reactive measures need to be developed.
For example, one preventive method~\cite{shan2023glaze} is to add toxic noises to the original images, designed to be visually negligible yet sufficient to mislead generative models with a substantial divergence in the semantic interpretation of the content. 
Regarding reactive measures, researchers are expected to develop more robust models to dissect the distribution differences between generative and real data, enabling discrimination and effective data governance~\cite{corvi2023detection, aghasanli2023interpretable}. 
}

\revise{\subsection{Bridging AI and Human Creativity in Commercial and Artistic Domains}}
\revise{
\noindent \textbf{Implications on commercial applications.}
While many users happily adopt easy-to-use platforms like Midjourney~\cite{midj23}, our comprehensive tool innovatively merges fine-tuning and generation into an intuitive workflow, offering enhanced customization for both creative and commercial endeavors.
For example, in the illustration industry, artists often receive commissions to create characters for animations or video games. 
Simply producing a single depiction of a character is insufficient. 
There's a need for multiple renderings of the same character from varied perspectives, in different attire and accessories, and set against diverse backdrops.
Creating more images of the same character from different angles, in other clothes and accessories, or even different environments is also important.
However, it is time-consuming for artists to finish all the drawings manually.
Our tool allows them to draw only a few images of their character and fine-tune their character model in subsequent AI generation for quick ideation and prototyping.
Furthermore, our tool adeptly assists in fine-tuning specific product models in commercial contexts like product advertising. This enables generating a wide range of product images, enhancing their appeal on online shopping platforms.
In addition, with the person's authorization, real human images can be used with our tool to support applications like virtual try-ons of fashion products or even make cross-generation photos or avatars showing elderly parents at their young age to facilitate intergenerational communication.   
}

\noindent \textbf{Bringing AI to the front of creative tools}. Traditionally, many users of generative AI treat the AI as a ready-made tool behind the scenes and only expect to use prompting on a simple interface to get the results, such as in popular tools like Midjourney~\cite{midj23} and Adobe Firefly~\cite{AdobeFire}. 
However, such a cooperation paradigm gives humans the illusion that they are controlling the AI even though AI is at the helm of the creative process.
Particularly, the AI algorithms are responsible for generating the content based on patterns and data they have learned from. 
Users are superficially guiding the AI, but the AI is doing the heavy lifting.
Our work has made an early step towards bringing generative AI to the foreground, stressing the importance of allowing users to control and customize the model. We achieve this goal by enabling users to intuitively teach AI new concepts and align them to user intents.

However, more work must advance a transparent and balanced co-creation paradigm between humans and AI.
Many previous explainable AI systems are expert-oriented, but for generative AI, which has attracted a much broader population of users, including artists, designers, and other AI novices, this poses new challenges to strike a balance between easy-to-use prompt interface and complex training and evaluation system.

\subsection{Limitations and Future Work}
\noindent \textbf{Considering extensibility}. 
Some expert users commented that although the baseline tool is designed with an engineering mindset and includes too many complex settings, it is built upon the Gradio~\cite{abid2019gradio} library, which is widely used for the quick implementation of rudimentary interfaces for models on Hugging Face, as P11 suggested. 
A notable advantage of such a tool is that it allows for easy extension with plug-ins developed by the community.
In comparison, our \textit{IntentTuner} is a specialized system that has not considered the easy addition of new functions by community users.
As the number of generative AI users continues to grow, so do their customization demands. This includes modifications not only to the fine-tuning process but also to the fine-tuning interface. 
In future work, we plan to incorporate extensible modules into our system that will allow users to add their own plug-in functions easily.

\noindent \textbf{Increasing the trust in LLM-assisted task}. 
We leverage a large language model to parse user intents.
However, despite the powerful reasoning ability of LLM, users sometimes feel they cannot fully trust its interpretation of their intentions.
The system should more intuitively display the intermediate results of LLM to allow for user adjustments instead of treating the LLM as a fail-proof panacea.
Users commented that our multi-modal intention input, which allows users to connect the language to visual concepts, can increase their trust to some degree because they feel the language interpretation is grounded by visual information.
In future work, we should develop more multi-modal visualization and interactions to allow users to view and refine AI interpretation of their intentions in fine-tuning.

\noindent \textbf{Achieving more sophisticated evaluation}.
The diffusion models still entail high uncertainty, as different random seed inputs can result in diverse generations.
Although our work incorporates human intentions into the evaluation, we have not addressed the uncertainty issue.
Moreover, domain expert users, like artists and designers, may have more sophisticated requirements for aligning aesthetic preferences.
For example, artists may want to preserve or change the color palette of the whole image instead of the hair color alone, or they may wish to control certain composition features.
These complex aesthetic features are challenging to describe in language and evaluate in the semantic space.
Particularly, the human preference metric used in our evaluation neglects the specific aesthetic aspects like color, composition, and stroke.
Future work needs to develop metrics and visualization methods that account for the uncertainty and concrete aesthetic aspects to improve the evaluation further.
For example, we could leverage some public datasets of fine-grained human-annotated aesthetic scores~\cite{murray2012ava} to train more comprehensive aesthetic evaluation metrics.

\section{Conclusion}
This study presents \textit{IntentTuner}, an intelligent framework to integrate human intentions in a novel but burgeoning AI-augmented creation task, which is customizing the text-to-image generative model with fine-tuning. 
The framework first allows users to articulate their fine-tuning intentions in natural multi-modal input, which is translated into structured intention specifications.
Then, the intention specifications guide the data augmentation to optimize the fine-tuning data in alignment with users' intentions.
Finally, \textit{IntentTuner} incorporates intent-aligned evaluation to help users evaluate the fine-tuning results based on their specific intentions instead of using generic metrics.
Based on the framework, we develop an integrated system that seamlessly combines the fine-tuning and generation functionalities to support a holistic and flexible workflow for text-to-image generation.
Application examples and a user study show that our framework and system can effectively help users reduce the trial and error workload and increase the intention alignment in fine-tuning. 
These show great potential in expanding the steerability and accessibility of generative AI to a broader group of users. 

\begin{acks}
\end{acks}

\bibliographystyle{ACM-Reference-Format}
\bibliography{reference}

\appendix
\section{Hyper-parameters setting of LoRA}
\label{appendix-lora}
\revise{ 
Table~\ref{tab:params} illustrated the default settings of model training hyperparameters. 
We choose  8-bit AdamW~\cite{dettmers20218} as the optimizer and use cosine annealing with warm restarts~\cite{loshchilov2016sgdr} as the learning rate (LR) scheduler.
Other hyper-parameters are preset based on the training domain translated from the user intents~(Sect. \ref{ssec:intent_specs}).
Stable diffusion consists of a U-net and a text encoder, with different learning rates during training.
The LoRA module has two essential parameters: Dimension and Alpha~\cite{hu2021lora}.
Dimension is the size of the low-rank update matrices, which determines the number of trainable parameters.
Alpha is a scaling factor that affects weight updates. 
Weight updates become more aggressive when $ \frac{alpha}{rank} $ is set to a higher value.

\begin{table}[h]\tiny
    \caption{Detailed setting of training hyperparameters}
    \resizebox{0.7\linewidth}{!}{
            \begin{tabular}{c|cccc}
    \toprule
    Domain   & U-net LR & Text encoder LR & Dimension & Alpha\\
    \midrule
    Painting   &  1e-4&1e-5 & 64 & 32\\
    Human portrait  & 1e-4 &5e-5 & 128 & 64\\
    2D character  &1e-4  &1e-5& 32 & 32  \\
    Product  &1e-4  &5e-5& 64 & 32\\
    \bottomrule
    \end{tabular}
    }
    \label{tab:params}
\end{table}
}

\end{document}